\documentclass[preprint,12pt]{elsarticle}

\usepackage{amssymb}
\usepackage{graphicx}% Include figure files
\usepackage{dcolumn}% Align table columns on decimal point
\usepackage{bm}% bold math
\usepackage{hyperref}
\hypersetup{colorlinks,allcolors=blue}
\usepackage{comment}
\usepackage{gensymb}
\usepackage{amsmath,amssymb,amsthm,mathtools,bm}
\usepackage{subfiles}
\usepackage{graphicx}
\usepackage{caption}    
\usepackage{subcaption}
\usepackage{float}
\usepackage{physics}
\usepackage{lipsum}
\usepackage[capitalize]{cleveref}
\usepackage[dvipsnames]{xcolor}

\usepackage[most]{tcolorbox}
\newtcolorbox{highlighted}{colback=yellow,coltext=black,breakable}

\journal{International Journal of Multiphase Flow}

\begin{document}

\begin{frontmatter}



\title{Quasi-steady evaporation of deformable liquid fuel droplets}


% \author{Meha Setiya \corref{cor1} \fnref{author}}
% \ead{setiyameha@vt.edu}
% \fntext[author]{Corresponding Author}
% \affiliation[cor1]{organization=Department of Mechanical Engineering, Virginia Tech %Department and Organization
%             % addressline={}, 
%             city={Blacksburg},
%             postcode={24061}, 
%             state={Virginia},
%             country={USA}}

%%original IJMF author style

% \author[author]{Meha Setiya \fnref{cor-author}}
% \ead{setiyameha@vt.edu}
% \fntext[author]{Corresponding Author}
% \affiliation[author]{%
%             organization={Department of Mechanical Engineering, Virginia Tech}, 
%             city={Blacksburg},
%             state={Virginia},
%             postcode={24061},
%             country={USA}}%

% \author[co-author]{John Palmore Jr.}
% \ead{palmore@vt.edu}
% \affiliation[co-author]{%
%             organization={Department of Mechanical Engineering, Virginia Tech},
%             city={Blacksburg},
%             state={Virginia},
%             postcode={24061},
%             country={USA}}%           

%% fix to get it worked with texlive2020

\author[author]{Meha Setiya\fnref{cor-author}}
\ead{setiyameha@vt.edu}
\fntext[author]{Corresponding Author}
%%\address[author]{Department of Mechanical Engineering, Virginia Tech, Blacksburg, Virginia, USA, 24061}

\author[author]{John Palmore Jr}
\ead{palmore@vt.edu}
\address[author]{Department of Mechanical Engineering, Virginia Tech, Blacksburg, Virginia, USA, 24061}

\begin{abstract}

This work covers the effect of droplet deformation on its evaporation rate under convective flow conditions. The evaporation behavior of a freely deforming droplet of single component jet fuel surrogate, n-decane, is investigated by varying Weber number ($We$) from $1-12$ and  Reynolds number ($Re$) from $25$ to $120$ under high-pressure environment. These studies utilize interface capturing Direct Numerical Simulation (DNS). To validate the accuracy of the solver, the results are compared against correlations by Abramzon and Sirignano (Int. Journal of Heat and Mass Transfer, 1989) and are found to be in good agreement with a maximum difference of $5 \%$. \textcolor{black}{A quasi-steady evaporation approach is implemented to simulate this problem.}
The results suggest a weak dependency of normalized total evaporation rate (${\dot{m}}_{ND}$) on Weber number at low $Re$ flow. However, a strong correlation is seen between the total evaporation rate and $We$ at high $Re$. $20 \%$ enhancement in ${\dot{m}}_{ND}$ is observed at $We=12$ (highly deformed shape) when compared to $We=1$ at $Re=120$.

In these cases, the distribution of local evaporation flux on the droplet is found to be proportional to its curvature up to the point of flow separation which agrees with low $Re$ theories on droplet evaporation by Tonini and Cossalli (International Journal of Heat and Mass Transfer 2013), Palmore (Journal of Heat Transfer 2022). 
Beyond the flow separation point, evaporation flux distribution depends on the boundary layer development and flow evolution downstream of the droplet. For highly deformed droplets, a larger wake region creates favorable fuel vapor gradients and promotes mixing in droplet wake, hence higher evaporation flux.

\end{abstract}


% \begin{highlights}
% \end{highlights}


\end{frontmatter}


\section{Introduction} \label{sec:Intro}

This work is inspired by spray combustion in gas turbines. The liquid fuel jet is injected into the hot compressed air in the combustion chamber where the jet breaks up into smaller fragments. The droplet size in the spray varies from a few microns to hundreds of microns \cite{sirignano_1999,book_ballal}. Within the spray, the overall evaporation behavior of the large droplets is different from the small droplets. For example, experimental studies such as \cite{VERWEY2017288} showed that the evaporation of small droplets or the droplets of a size approaching the Kolmogorov length scale stay unaffected by the presence of turbulence in a zero mean velocity flow. However, the evaporation of the large droplets is enhanced by the turbulent environment even in zero mean velocity. Moreover, in real sprays, although the larger droplets are fewer in the count, their volume fraction contributes to the majority of the spray volume. This is because the volume of a droplet is proportional to the cube of its diameter. \textcolor{black}{Therefore, the droplet size plays an important role in the atomization process and further in combustion \cite{book_lefebvre}.}
With an emphasis on large droplets, this work focuses on analyzing the effect of droplet shapes on their evaporation.

Due to the presence of turbulent and convective flow conditions inside combustion chamber \cite{CHO200914, Rauch2012}, an imbalance of inertial and surface tension exists. Hence, droplets can deform from a spherical shape to a non-spherical shape \cite{Mashayek2001}. This deformation can be characterized by a non-dimensional number called ``Weber number" \cite{loth2006} defined as,
\begin{equation}
    We= \frac{\rho_G U_{rel}^2 d_0}{\sigma}
    \label{eq:Weber}
\end{equation}
Here, $\rho_G$ is gas density, \textcolor{black}{$U_{rel}= (U_{in}-U_{droplet})$ is the relative gas velocity with respect to droplet velocity}, $d_0$ is the initial diameter of the droplet and $\sigma$ is the surface tension of the liquid droplet. As Weber number governs the droplet shape, small Weber numbers ($We<1$) lead to nearly spherical shapes. Whereas, large Weber numbers lead to more deformed shapes. Therefore, for the same liquid properties, small droplets will have lower Weber numbers, hence nearly spherical shapes. Whereas, larger droplets will have higher Weber numbers, hence complex deformed shapes.

A great amount of experimental, as well as numerical studies on droplet evaporation and combustion are available with the assumption of droplets as a spherical ``non-deforming" particles under convective flow conditions \cite{KOTAKE1969595, RAGHAVAN2005, KUMAGAI1957726}. However, there is a dearth of theoretical and analytical literature on the evaporation and combustion of ``deformable" droplets. Before diving into the evaporation of deformable droplets, this paper reviews major results on the evaporation of spherical droplets first.
\textcolor{black}{It is important to highlight that the term evaporation rate is somewhat ambiguous. In the combustion community\cite{turns2000}, ``evaporation rate'' is widely used to refer to the net rate of change in mass of the droplet. However, in other contexts, the term ``evaporation rate'' is also used to describe the net rate of change in mass taken per unit surface area. In an effort to remove such ambiguity, the authors introduce the terminology of ``total evaporation rate (net rate of change in mass)" and ``evaporation flux (net change in mass taken per unit surface area)" which will be used throughout this work.} 

A widely common and fundamental approach in this field is to study quasi-steady droplet evaporation under quiescent ambient conditions. The result of such studies is the $d^2$ law which states that the total evaporation rate of a spherical droplet is proportional to its instantaneous diameter. Moreover, the droplet square diameter decreases linearly with time under such ambient conditions \cite{sirignano_1999,turns2000}.  

An analytical study on droplet evaporation under convective flow by Law et. al \cite{LAW1977605} simplifies the flow over droplet by representing it as the flow of a gas stream over a liquid fuel stream. In the pressurized environment, the liquid surface velocities are significant when compared to free stream velocity. The higher surface velocity affects the internal motion in liquid which affects the evaporation rate. It was also shown that an increase in evaporation rate leads to a reduction in the interface velocity due to the thickening of the boundary layer around the liquid surface.

Such alteration in the gas phase boundary layer was further elaborated in a work by Renksizbulut and Yuen \cite{Renksiz1983} who studied the effect of the composition of fuel-air and gas flow near the droplet on its evaporation rate.  
These numerical studies were performed at Reynolds numbers ranging from $Re=10$ to $Re=100$ for various liquid droplets and a solid sphere. Here, \textcolor{black}{Reynolds number based on the initial droplet diameter $d_0$ can be defined as $Re=\frac{\rho_GU_{rel}d_0}{\mu_G}$}.
The results of this study revealed that decreasing the droplet surface temperature does not directly increase the Nusselt number (hence the evaporation rate) due to the presence of a steeper temperature gradient. The cold boundary layer of gas affects the evaporation rate via its effects on the thermal conductivity of the gas.

A work by Haywood et. al. \cite{Haywood1989} focused on the variable thermophysical properties of gas and transient processes in gas and liquid phase involved in droplet evaporation. The transient processes in the liquid phase were receding droplet size, fluid motion inside the droplet, and liquid heating. 
For inflow at $Re=100$, an increase in the Nusselt number was seen in the early lifetime of the droplet due to the beginning of the surface blowing. Once, steady evaporation establishes, Nusselt number variation flattens with time. The angular variations in droplet surface temperature and fuel vapor mass fraction were more prominent at the early stage of evaporation due to strong sensitivity to vapor pressure. 
Moreover, due to the unsteady nature of liquid motion, the liquid temperature rises quickly up to the droplet surface temperature as time progresses and liquid heating can be considered quasi-steady.

Another common assumption in evaporation studies is neglecting liquid heating due to radiation. A recent work by Merino et. al. \cite{millan2021} quantified the effect of radiation using order of magnitude analysis. They studied evaporation of a single droplet in an inert stagnant environment under micro-gravity conditions. The authors identified the relevant timescales involved in evaporation. It was found that for larger droplets ($d>0.5 mm$) in high ambient temperature, the timescales for radiation heating and conduction heating are of the same order. Hence, $d^2$ trend can deviate significantly from the expected linear profile.

A recent comprehensive numerical work by Ray et. al. \cite{RAY2019294} quantified the effect of pressure, temperature, convection and initial-liquid phase composition on droplet evaporation. The droplet shape was considered to be nearly spherical as Weber number was $We \approx 0.3$ for the given conditions. Results showed faster evaporation rates at higher pressure due to net effect of decrease in gas diffusivity and surface tension, and increased latent heat. Higher ambient temperature and higher convective velocity favor the reduction in droplet lifetime.

The previous literature includes a wide variety of approaches used to study the evaporation of spherical droplets. Several major results exist demonstrating the effect of various environmental factors on their evaporation. However, much less work has directly studied the effect of deformation on evaporation. Haywood et al. \cite{haywood1994} studied the dynamics of evaporating droplets in convection using a 2D (axisymmetric) finite volume code. The code used a dynamic body-fitted mesh to solve for the droplet deformation and interface recession. However, their study focused on the effect of evaporation on droplet drag, and results for the effect of deformation on the evaporation rate were not studied in detail. 

Mashayek \cite{Mashayek2001} studied the effect of shape on droplet evaporation in quiescent flow using a finite element code. The work chose to impose droplet shape using arbitrary linear combinations of spherical harmonics. The work showed that evaporation enhances due to deformation, and it suggested that local evaporation could be related to the interface curvature.

Schlottke and Weigland \cite{SCHLOTTKE20085215} developed one of the first direct numerical simulation solvers for droplet evaporation in three dimensions. They validate the solver with correlations by Ranz and Marshall \cite{ranz1952}, Kumala et al. \cite{KULMALA1995}, Renksizbulut et al \cite{Bulut1991}. Their results clearly indicated the non-uniform evaporation that occurs over the surface of the droplet, and the corresponding non-uniformity of vapor in the wake. 

A theoretical study by Tonini and Cossali \cite{tonini2013,TONINI2016} derived analytical solution for the steady state evaporation of ellipsoidal droplets under quiescent flow assuming constant density in both phases. To take into account of droplet non-sphericity on its evaporation, analytical expressions for evaporation rate and evaporation flux for ellipsoidal droplets (prolate and oblate shapes) were developed which generalize to the spherical droplet as well. The deformation parameter is $\epsilon= a_r/a_z$, where $a_r$ is the radial spheroid axis and $a_z$ is axial spheroid axis.  Another variable $\beta$ is ratio of surface area of spheroid ($A_{sphd}$) to the spherical droplet ($A_{sph}$), for the same liquid volume $\beta=A_{sphd}/A_{sph}$.
Results showed for the same volume and $\beta$, prolate droplet ($\epsilon<1$) has higher total evaporation rate in comparison to oblate shape ($\epsilon>1$). Furthermore, the contours of vapor concentration revealed the relation between local vapor flux and the surface curvature is related to the fourth root of local Gaussian curvature of the droplet. This suggested that the maximum vapor flux is seen at the higher curvature region and the low flux is low curvature or flatter regions of the droplet.

The idea of non-spherical droplet evaporation was further expanded by Palmore  \cite{palmore_JHT_2022} using a perturbation theory in his recent work. The work looked at evaporation and combustion of non-spherical droplets. The use of more realistic physics-based droplet shapes \cite{taylor_acrivos_1964} than the ellipsoid is one of the novelties in his work. The assumption for this work were, the droplet at its saturation temperature, large density and viscosity ratio between liquid and gas, single component fuel and unity lewis number. The theories were established for a quiescent flow conditions. For a fixed $Re$ flow, $We$ was selected as a perturbation parameter. 
The results agreed with the findings by Mashyek \cite{Mashayek2001} and Tonini and Cossali \cite{tonini2013} that the local evaporation flux relates to the local interface curvature. Moreover, higher total evaporation rate for prolate shapes than oblate shapes within the limit of smaller deformations was consistent with \cite{tonini2013}, \cite{TONINI2016}.
The author also substantiated that although the total surface area of non-spherical droplet is higher than the spherical droplet, the total evaporation rate droplet does not scale by the total surface area ratio of non-spherical to a spherical droplet. In other words, it is the local evaporation flux that is affected by deformation. 

It is noted again that the theoretical analysis by Tonini and Cossalli \cite{tonini2013,TONINI2016}, Palmore \cite{palmore_JHT_2022} as well as the numerical simulations by Mashayek \cite{Mashayek2001} were performed under quiescent flow. Hence, a true quantification of the effect of deformation on the evaporation of freely-deforming droplets under planar convective flow is yet unknown.
Therefore, the goal of this work is to gain insights about the interaction of droplet deformation and its evaporation rate under convective flow using DNS. 

This work is organized as follows: \cref{sec:method} includes the details about the flow solver. \cref{sec:setup} discusses the numerical setup and boundary conditions. Subsequently, \cref{sec:validation} includes the grid independence study and validation of solver with Abramzon and Sirignano analytical correlations. Finally \cref{sec:shape_evpaoration} demonstrates the main study on the influence of the deformation and planar flow on evaporation. The paper concludes with \cref{sec:conclustion}.

%------------------------------------------------------------
\section{Methodology} \label{sec:method}
This study utilizes an in-house framework for interface-capturing direct numerical simulations of evaporating multiphase flows~\cite{palmore_2019}. It is built upon the conservative finite volume framework of~\cite{nga}. The details of numerical methods and algorithm can be found in~\cite{palmore_2019}, and validation study in~\cite{palmore_iclass}. A brief description of flow solver is mentioned in this section.  Subscripts $L$, $G$, $V$ refer to liquid, gas and vapor phase respectively throughout this work.
%_____________________________________________________________________________
\subsection{Flow solver} \label{flow_solver}
The solver solves the dynamics of flow using first principles i.e. conservation of mass, momentum and energy. These can be described in their mathematical form as follows. 

The conservation of momentum for a Newtonian fluid can be written as,
\begin{subequations}
\begin{align}
  \frac{\partial\left(\rho\bm u\right)}{\partial t}+\nabla\cdot\left(\rho\bm u\otimes\bm u\right)=-\nabla p+\nabla\cdot\left(\mu\bm S\right), \text{ where} \label{eq:NS1}  \\
  \bm S=\left(\nabla\bm u+{\nabla\bm u}^\top-\frac{2}{3}\nabla\cdot\bm u\right), \label{eq:NS2}
\end{align}
\end{subequations}
Here, $\rho$ and $\mu$ are the fluid density and dynamic viscosity; $\bm u$ is the velocity; and $p$ is the pressure.

\cref{eq:NS1} must be coupled with the continuity equation for incompressible flow given as,
\begin{align}
\nabla\cdot \bm u=0
%   \frac{\partial\rho}{\partial t}+\nabla\cdot\left(\rho\bm u\right)=0,\label{eq:cont}
\end{align}
The pressure is computed from the incompressibility constraint for the velocity field. It is solved using pressure Poisson equation using Ghost Fluid Method (GFM) by Liu et. al. \cite{LIU2000151}. 

With the assumption of low-Mach number flow, neglecting the heat due to viscous dissipation and energy generation, the conservation of energy in the liquid phase can be simplified as,
\begin{equation}
\frac{\partial (\rho_L\ c_{p,L}\ T_L)}{\partial t} +\nabla\cdot(\rho_L\ c_{p,L}\ T_L \ \bm u) =  \nabla \cdot(\rho_L\ c_{p,L}\lambda_L \nabla T_L) \label{eq:liq_energy}
\end{equation}

Here $\lambda_L = \ k_L / (\rho_L \ c_{p,L})$, known as thermal diffusivity. 
The energy equation for gas phase can be written in the same fashion.

The transport of fuel vapor specie in the gas phase can be specified as,
\begin{equation}
\frac{\partial (\rho Y) }{\partial t} +\nabla \cdot( \rho Y \bm u) =  \nabla .( \rho D \ \nabla Y)  \label{eq:specie}
\end{equation}
Here $Y$ is  species mass fraction of fuel vapor and $D$ is diffusion coefficient.
The scalar quantities, temperature and chemical species are solved using two field approach by Ma and Bothe in \cite{MA2013552}.

Finally, for constant thermophysical properties, the transport equation for liquid volume fraction ($\alpha$) is given as,
\begin{equation}
\frac{\partial \alpha }{\partial t} +\nabla \cdot(\alpha \bm u) = - \frac{\dot{m''}}{\rho_L} \delta
\label{eq:alpha}
\end{equation}
\textcolor{black}{The liquid mass fraction $\alpha$ is defined such that it is  1 in a purely liquid cell and 0 in a purely gas cell. This solver uses a directionally-unsplit geometric volume of fluid method to evolve the $\alpha$ field. The method utilizes a sharp interface representation with explicit interface reconstruction with the Piecewise Linear Interface Calculation (PLIC) method. As a consequence of this approach, the interface always lies within only one computational cell. Utilizing PLIC all geometric information about the interface location and normal vector is generated. The interface curvature is computed based on a least-squared distance field generated from the PLIC reconstruction. The readers are referred to \cite{palmore_thesis} for more details of numerical algorithms used for interface reconstruction.}

The surface area density $\delta$ for any arbitrary computational cell $\Omega$ is defined as,
\begin{align}
     \delta= \int_{\Gamma \cap \Omega} dS \bigg / \int_\Omega  dV \label{eq:delta} 
\end{align}
Here, $\Gamma$ refers to Liquid-Gas interface.
And, $\dot{m''}$ is local evaporation rate per unit area.

%_____________________________________________________________________________
\subsection{Matching conditions at the liquid-gas interface}
In order to ensure the conservation of mass, momentum and  chemical species at the interface, matching conditions at the liquid-gas interface need to be specified. The condition for mass conservation at interface implies,

\begin{equation} \label{eq:vel_jump}
[\bm n.\bm u]_\Gamma = \dot {m''} \bigg[\frac{1}{\rho}\bigg]_\Gamma
\end{equation}

Here the bracket notation is used for the jump across the interface. For any variable $\phi$, $[\phi]_\Gamma= \phi_G-\phi_L$, $\dot{m''}$ is local evaporation flux.

The matching condition for momentum is,
\begin{equation}\label{eq:press_jump}
[p]_\Gamma = -\sigma \kappa - {\dot{m''}^2}\bigg[\frac{1}{\rho}\bigg]_\Gamma
\end{equation}
where $\sigma$ is surface tension and $\kappa $ is local curvature of the interface and is defined to be positive when the liquid is locally convex. 

The energy balance due to evaporation at the interface can be written as, 
\begin{equation}
\frac {[\bm n\cdot k \nabla T]_\Gamma}{\ L_v} = \dot{m''}
\label{eq:gradT}
\end{equation}
where $L_v$ is enthalpy of vaporization. Lastly, the species conservation requires
\begin{equation}
\dot{m''}Y - \bm n \cdot \ \rho_G \ D \ \nabla Y = \dot{m''}
\label{eq:gradY}
\end{equation}
Here, $D$ is the diffusivity of vapor in gas phase. \cref{eq:gradY} is solved only in gas phase, hence simplified. 

As a general approach, the thermodynamic equilibrium at the interface is maintained using the Clausius-Clapeyron relation \cref{eq:CC_rel} and \cref{eq:moletoY}.
\begin{equation}
    X= e^{\frac{-L_vM_v}{R}\big(\frac{1}{T_\Gamma} - \frac{1}{T_{sat}} \big)}
    \label{eq:CC_rel}
\end{equation}
\begin{equation}
    Y= \frac{XW_v}{XW_v+(1-X)W_G}
    \label{eq:moletoY}
\end{equation}
Here, $W_v$ and $W_G$ refer to molecular weight of the fuel vapor and ambient gas. $X$ refers to mole fraction and $Y$ refers to mass fraction of fuel vapor. 
For a given $\dot{m''}$, \cref{eq:CC_rel}, \cref{eq:gradT} and \cref{eq:gradY} must be satisfied simultaneously and the strategy for doing so is discussed in \cite{palmore_2019}. The presented studies consider the droplet to be at its saturation conditions.

\subsection{Quasi-steady evaporation}
\textcolor{black}{
This article deals with quasi-steady droplet evaporation. From the thermodynamics point of view, a process is called quasi-steady when the system is slowly changing its state (i.e. pressure, temperature, entropy) while staying in continuous equilibrium with the surrounding. The quasi-steady assumption is very common within the literature of droplet evaporation and combustion studies, and many of the previously mentioned studies actually were in this regime. The $d^2$ law, a fundamental observation in droplet evaporation and combustion, is also derived under quasi-steady configurations.}

\textcolor{black}{In this computational code, the quasi-steady approach is implemented by fixing the liquid volume as constant throughout the simulation. Numerically, this is achieved by ignoring the evaporation flux term in \cref{eq:alpha}. However the term is not ignored in other relations (\cref{eq:vel_jump,eq:press_jump,eq:gradY,eq:gradT}). This approach is somewhat analogous to the experimental studies on droplet evaporation using porous spheres \cite{Renksizbulut_porous, YUEN1978537}. In that context, a liquid layer fully coats the outside of the sphere, and it evaporates slowly. Each sphere is fed with fluid through a tube such that the liquid volume remains constant throughout the study. This numerical approach for quasi-steady evaporation can especially be helpful in developing correlations of total evaporation rate as a function of initial $Re$ and $We$. As in this case, the values for these quantities are constant in time.
}

%------------------------------------------------------------
\section{Numerical Setup} \label{sec:setup}
The numerical setup for this study includes a single deformable droplet of diameter $d_0=100\ \mu m$ evaporating in a hot and pressurized incoming flow at $P_\infty=20 \ atm$ \cite{wu2010} and $T_\infty=750 \ K$. These conditions were selected to mimic environment of a gas turbine combustor. As a simplification, the liquid droplet is at its boiling temperature $T_\Gamma = 615 \ K$. Hence, the internal heating of the droplet is not considered. Constant thermophysical properties of air and liquid fuel at these ambient conditions are listed in  \cref{table:1}. 
% Surface tension ($\sigma$) is not listed in \cref{table:1} as \textcolor{black}{this value is modified to vary the Weber number based on \cref{eq:Weber}} .
\begin{table} [H]
\begin{center}
\begin{tabular}{ |p{4.2cm}|p{2.4cm}|p{2.4cm}|p{2.4cm}| } 
\hline
Property & Units& Air & Fuel\\ 
\hline
\hline
Density $\rho$ & $kg/m^3$ & 9.41 & 300 \\ %%299.99 \\ 
Viscosity $\mu$ & $kg/(m\cdot s)$ &$3.48\times10^{-5}$ &$4.00\times10^{-4}$\\ 
Specific heat $c_P$ & $J/(kg\cdot K)$& 1086 & 13220 \\ %% 1086.11, 13224
Thermal conductivity $k$ & $W/(m\cdot K)$ & 0.054 &0.079\\ 
% Surface tension $\sigma$ & $N/m$ & --- &0.010\\ 
Latent heat of vaporization $L_v$ & $J/kg$& --- &$4.81\times 10^4$\\ 
Boiling temperature $T_{boil}$ & $K$ & --- &615.05 \\ 
\hline
\end{tabular}
\end{center}
\caption{Fluid properties at $P_\infty=20 \ atm, T_{boil}=615 K, T_G=750K$ \cite{nist_webbook}}
\label{table:1}
\end{table}

\begin{figure} [h!]
    \centering
        \includegraphics[width=0.7 \linewidth]{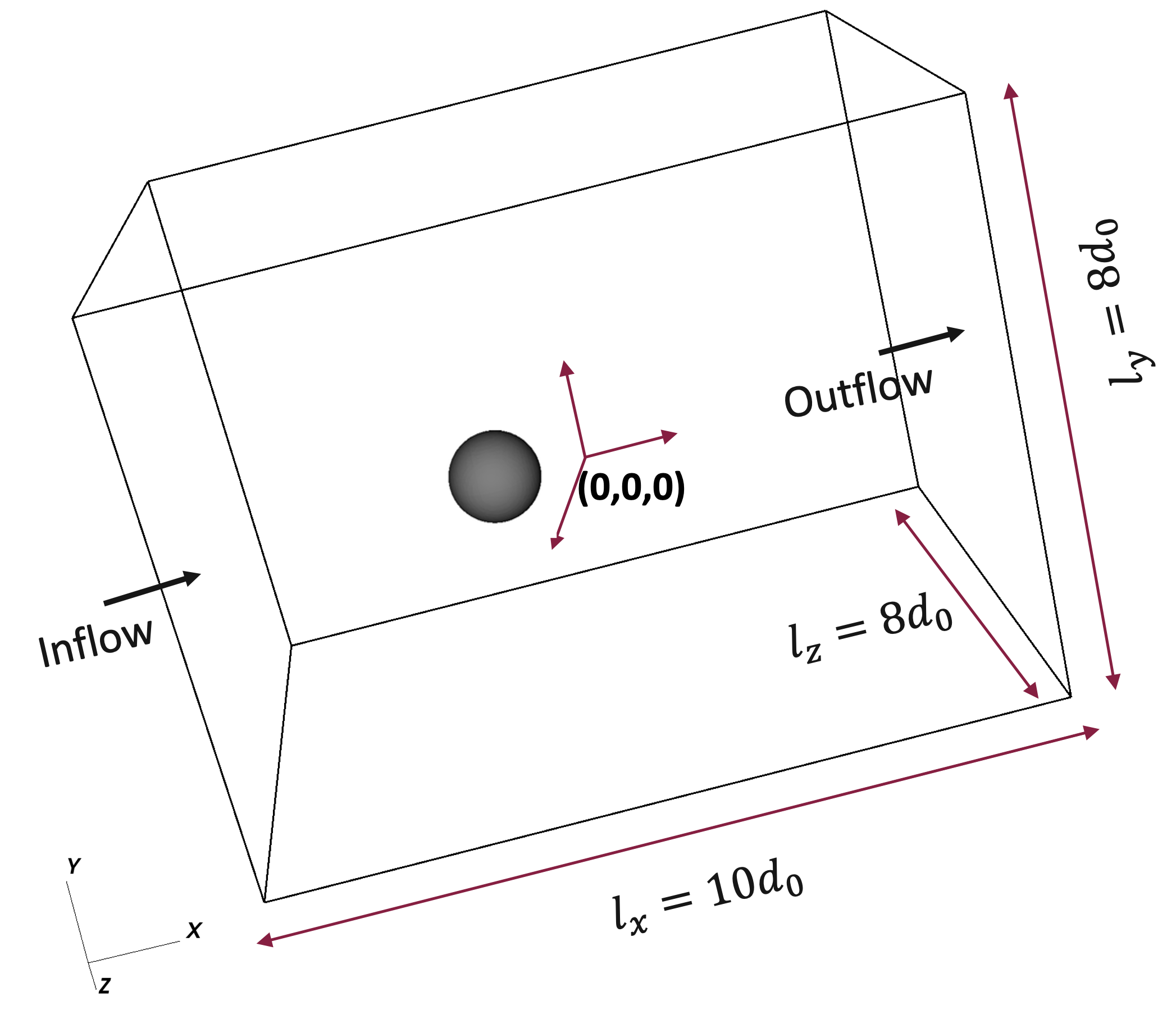}
    \caption{A sketch of numerical setup and boundaries}
    \label{fig:ch_1domain_3d}
\end{figure}

The domain is of size  $10d_0 \times 8d_0 \times 8d_0$ and the droplet is located at ($-d_0,0,0$) with respect to origin ($0,0,0$) as shown in  \cref{fig:ch_1domain_3d}. This domain size and the location of droplet center is selected such a way that the boundaries do not influence the flow field around the droplet.

A uniform structured grid is used for this work. As depicted in  \cref{fig:ch_1domain_3d}, a fixed inflow boundary condition is applied on the negative x-face of the domain. To keep the droplet at its predetermined location, a method is deployed which mimics the flow over droplet as falling with its terminal velocity ($U_T$). 
This method has been tested and detailed in our previous works \cite{yushu_scitech, setiya2020}. \textcolor{black}{This results in nearly constant $U_{rel}$}. An outflow boundary condition is applied on the positive x-face of the domain. The remaining faces are treated as periodic boundaries. \textcolor{black}{The domain is initialized with initial conditions obtained using $d^2$ law.}
This numerical setup is kept same for all the studies presented in this work. 

\textcolor{black}{It is important to note that because quasi-steady conditions are sought, data are analyzed after an initial transient period has ended. This transient period exists because droplets tend to oscillate when exposed to the flow. It takes some time for the internal forces on a deformed curved surface (surface tension and viscous forces) to balance with external aerodynamic pressure to reach a stable droplet shape \cite{SCHLOTTKE20085215}. The time to reach this shape is roughly, proportional to the capillary time scale which will be introduced in~\cref{sec:result}.}

%------------------------------------------------------------
\section{Results and Discussion}\label{sec:result}
\subsection{Grid Independence Study}\label{sec:validation}

A grid convergence study is performed to find an optimum grid resolution required to capture the dynamics for the given problem. This is done by varying the grid length $dx=  12.5, \ 6.25,\ 4.17, \ 3.13, \ 2.08  \ \mu  m$. This corresponds to the number of cells per initial droplet diameter $N_{d_0} = 8,\ 16,\ 24,\ 32, \ 48$. \textcolor{black}{The flow conditions $Re=120$ \cite{taneda1956}, $We=12$ are selected for this study.  This Weber number is chosen to be below the critical Weber number for the onset of breakup which is usually 12 ~\cite{loth2006,KEKESI20141,hinze1955}. However, it is important to note that the droplet breakup does not solely depend upon Weber number. Other conditions such as density ratio, viscosity ratio, Ohnesorge number, flow velocity and type of flow around it affect the breakup time and the breakup mode \cite{PILCH1987741, Villermaux2009}. However, for a wide range of conditions relevant to liquid droplets in the air, $We \lesssim 12$, is sufficiently predictive for non-breakup.}

\begin{figure} [h!]
    \centering
        \includegraphics[width=0.6 \linewidth]{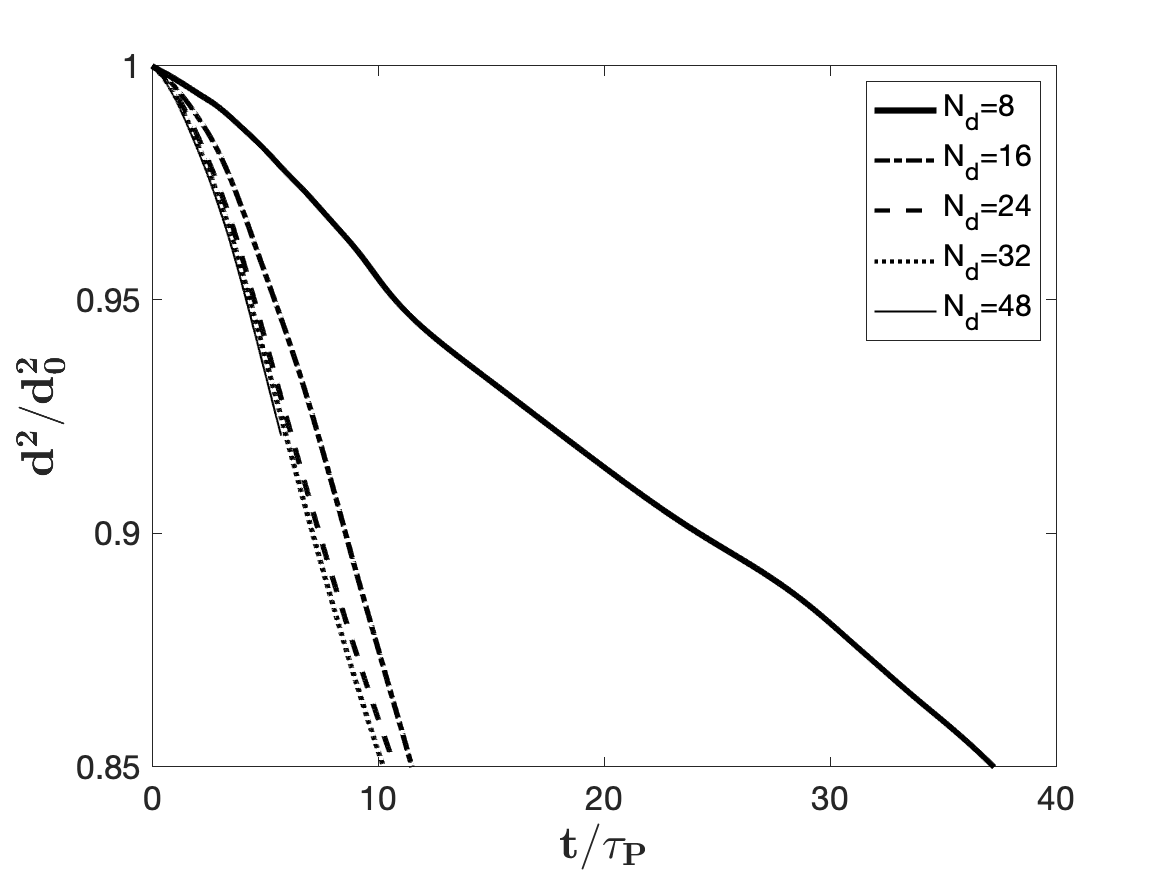}
    \caption{$d^2/d_0^2$ vs normalized time for various grids }
    \label{fig:d2}
\end{figure}

The results from the grid independence study are plotted in terms of a non-dimensional parameter $d^2/d_0^2$ with non-dimensional time ($t/\tau_p$) as shown in \cref{fig:d2}. 
\begin{equation}
\tau_p = \sqrt{\frac{\rho_L +\rho_G}{\sigma}} {\bigg(\frac{d_0}{2\pi}}\bigg)^{3/2}
\end{equation}
\textcolor{black}{ $\tau_p$ is capillary timescale \cite{Sussman2009}}. It is noted that the slope of each curve are approximately constant after an initial transient period. The results also show that as $N_{d_0}$ increases, the magnitude of the slope of droplet decay increases. \textcolor{black}{Moreover, the slope of curve for $N_{d_0}=24$ is observed to be nearly the same as $N_{d_0}=32$ and $N_{d_0}=48$.}
% No grid dependent droplet break-up is noticed for other grids except at $N_{d_0}= 8,\ 16$ } 

\begin{figure} [h!]
    \centering
        \includegraphics[width=0.6 \linewidth]{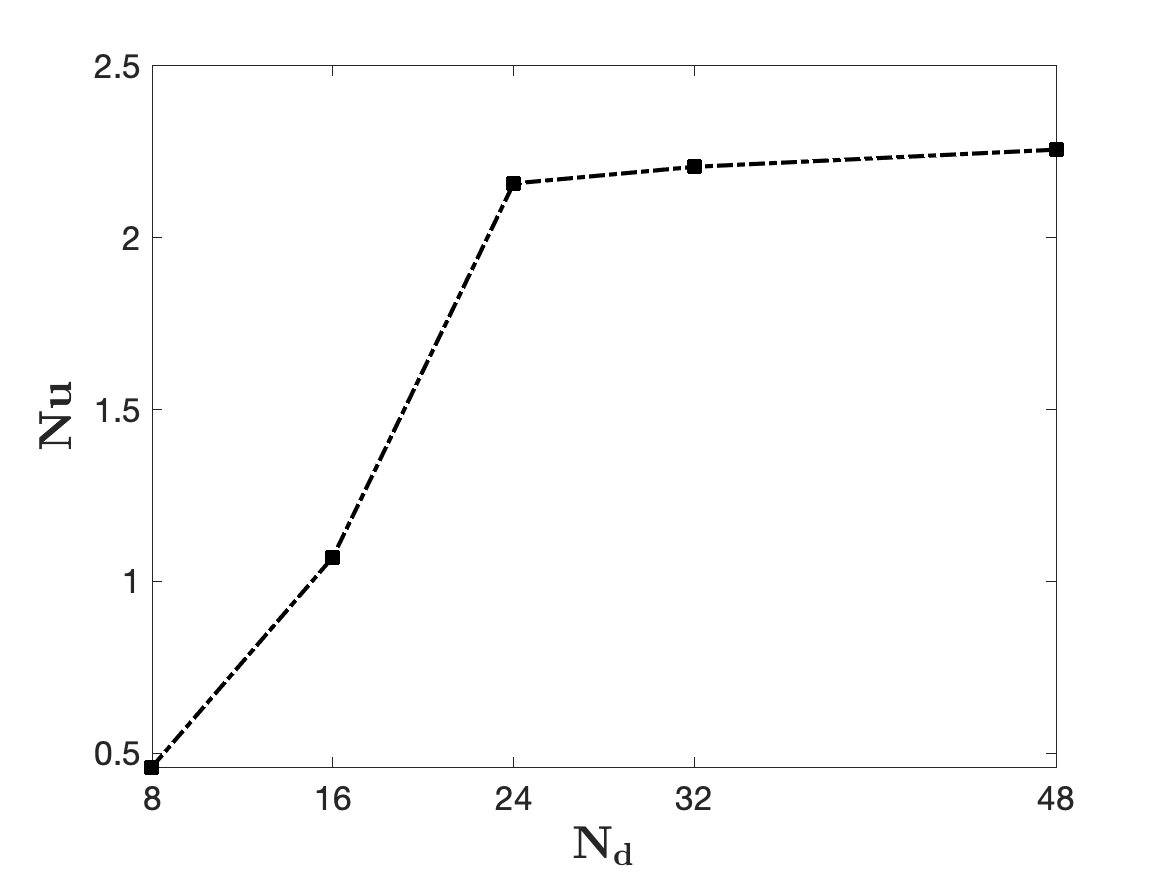}
    \caption{Nusselt Number ($Nu$) with cells per droplet diameter ($N_{d_0}$)}
    \label{fig:grid_convergence}
\end{figure}

Various grids are further compared in terms of Nusselt number $Nu$, which can be expressed as,
\begin{equation}
 Nu = -\frac{2r}{(T_\Gamma - T_\infty)} (\bm n_\Gamma \cdot \nabla T)   
\end{equation}
Here $r$ is the radius of droplet at that instant and $\bm n_\Gamma$ is the normal vector at the liquid-gas interface. \cref{fig:grid_convergence} shows the surface average $Nu$ for each grid. \textcolor{black}{It is noted that unphysical droplet breakup was observed for $N_{d_0}=8$ and $16$, however, this does not appear at $N_{d_0}= 24$. Further refinement beyond $N_{d_0}= 24$ does not affect the breakup behavior of the solution.} As the grid resolution is increasing, the gradients of temperature are captured more finely. Hence, Nusselt number is observed to be increasing. \textcolor{black}{In case of $N_{d_0}=24$, Nusselt number is found to be only $2.2\%$ and $4.5\%$ lower with respect to $N_{d_0}=32$ and $N_{d_0}=48$ respectively. Hence, $N_{d_0}=24$ is found to be computationally appropriate for this work.}

%%%%%%%%%%%%%%%%%%%%%%%%%%%%%%%%%%%%%%%%%%%
\subsubsection{Validation}
This section discusses the validation of the numerical results against a semi-analytical theory of droplet evaporation developed by Abramzon and Sirignano \cite{abramzon1989}. This theory represents evaporation as rapid heat and mass transfer in a thin film over the droplet. The film theory predicts the Nusselt number as a function of \textcolor{black}{Spalding heat transfer number ($B = c_{p_g} \ (T_\infty - T_\Gamma/h_{fg}$)}, Reynolds number ($Re$) and \textcolor{black}{Prandtl number ($Pr = \mu c_p / k$)}.
The film theory uses a correlation for the mass transfer in non-evaporating flows and further modifies it for to account of the film due to evaporation. 
Nusselt number in a non-evaporating flows is denoted as ${Nu}_0$ and Nusselt number for evaporating flows is denoted as ($Nu$). After correction in ${Nu}_0$ due to evaporation, $Nu$ can be written as,
\begin{equation}
    Nu = 2 \bigg(\frac{\ln(1+B)}{B}\bigg) + ({Nu}_0-2)(1+B)^{-0.7}
    \label{eq:nu_ans}
\end{equation}
The study in \cite{abramzon1989} includes two expression for ${Nu}_0$. The first correlations for ${Nu}_0$ by Clift et al. \cite{Clift1978} (CEA) is as follows,
\begin{equation}
    {Nu}_{0,CEA}= 1+ (1+Re \ Pr)^{1/3} f(Re)
    \label{eq:clift}
\end{equation}
Where, $f(Re)=1 $ when $Re \le 1$ and  $f(Re)=Re^{0.077} $ when $Re>1$ in  \cref{eq:clift}.

\begin{figure}[h!]
   \centering
   \includegraphics[width=0.6 \linewidth]{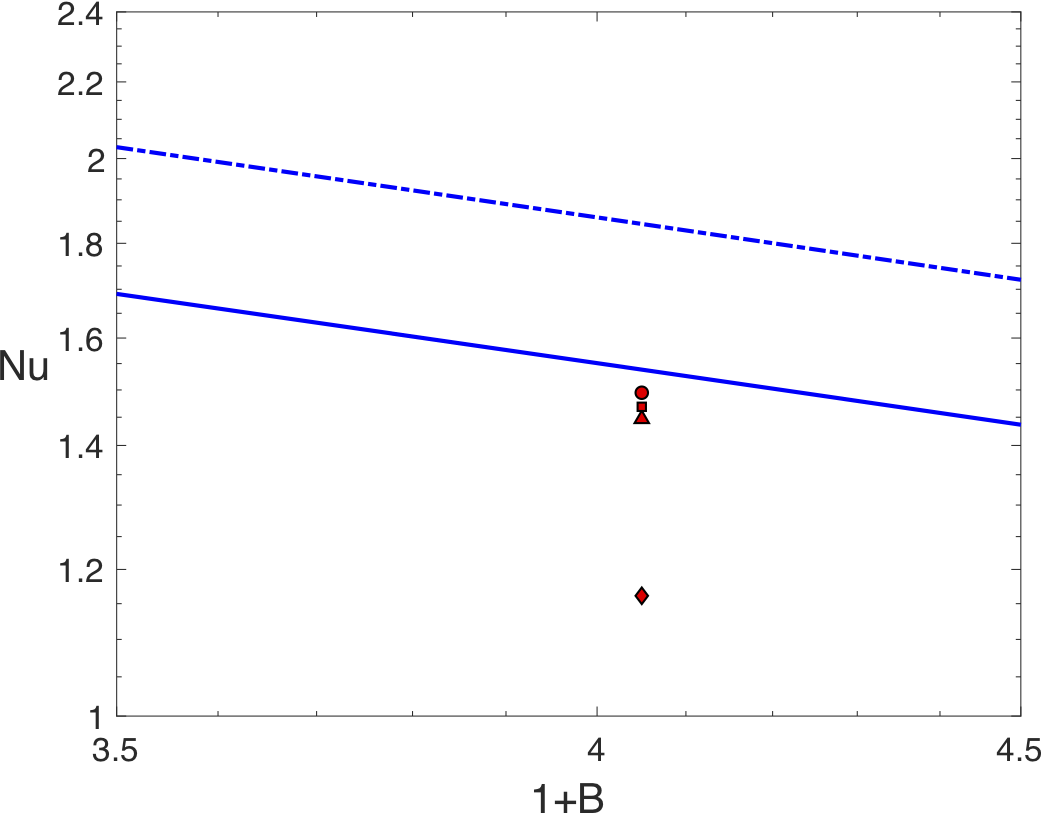} 
   \caption{Nusselt number vs Spalding heat transfer number. \textit{Blue solid line: Clift et al., Blue Dash line: Ranz and Marshall, Red markers: Numerical Results [Diamond marker: $N_{d_0}=8$, Triangle marker: $N_{d_0}=16$, Rectangular marker: $N_{d_0}=24$, Circular marker: $N_{d_0}=32$] }}
   \label{fig:AC_Re_25_we_1}
\end{figure}

The second correlation for ${Nu}_0$ by Ranz and Marshall \cite{ranz1952} (RM) is as follows,
\begin{equation}
    {Nu}_{0,RM}= 2+ 0.552 Re^{1/2} Pr^{1/3}
    \label{eq:Ranz}
\end{equation}
The correlation from CEA is known to be more accurate, particularly at small Reynolds numbers. However, the RM correlation appears to be more widely used in practice. The modified Nusselt number for evaporating flow ($Nu$) using these two correlations is shown in  \cref{fig:AC_Re_25_we_1}. The plot shows the comparison of Nusselt number for an evaporating sphere ($Nu$) using Clift et. al. (solid blue line) and  Ranz and Marshall (blue dot-dahsed line) at various Spalding heat transfer numbers.

Now, in order to test the accuracy of solver against \textcolor{black}{the empirical correlations}, the numerical results for the following conditions are overlaid on it. The flow conditions $Re=25$ \cite{taneda1956}, $We=1$ \cite{loth2006} are selected for this study. These conditions are chosen to best match with \textcolor{black}{the empirical correlations} which deal with spherical droplets with fully attached boundary layer. This is done by varying the grid length $dx= 12.5, \ 6.25, \ 4.17, \ 3.13 \ \mu  m$. This corresponds to number of cells per initial droplet diameter $N_{d_0} = 8,\ 16, \ 24, \ 32$.

The numerical results are shown here as markers in red with outline of black color. For a particular Spalding number $B=3.05$, an increasing trend in Nusselt number is seen with the grid refinement. Additionally, Nusselt number from the numerical simulation move towards the correlation. 
\begin{table} [h!]
\begin{center}
\begin{tabular}{ |p{1cm}|p{2cm}|p{2cm}|} 
\hline
\bm{$N_{d_0}$} & \bm{$Nu$} & \bm{$\%$} \textbf{error}\\ 
\hline
\hline
8 & 1.161 & 24.49 \\
\hline
16 & 1.446 & 5.93 \\
\hline
24 & 1.468 & 4.50\\
\hline
32 & 1.494 & 2.82 \\
\hline
\end{tabular}
\end{center}
\caption{$\%$ error in Nusselt Number ($Nu$)}
\label{table:Nu}
\end{table}
The difference between the simulation results and the CEA correlation (\cref{eq:clift}) is much less than that of the CEA correlation to the RM correlation (\cref{eq:Ranz}). This suggests that the simulation results are within a reasonable band of error against the true solution. The
$\%$ error with respect to empirical correlations from Clift et. al. is tabulated in \cref{table:Nu}.  Less than $5\%$ error is seen between the simulation results and the CEA correlation for mesh size greater than $N_{d_0}=24$.
Therefore, based on these analysis, $N_{d_0}=24$ is found to be an appropriate grid size in terms of computational expense as well as accuracy for further work.

%---------------------------------------------------------------------------------------

\subsection{Effect of droplet shape on evaporation}\label{sec:shape_evpaoration}

After finding the appropriate grid and validating the results with the \textcolor{black}{empirical correlations}, this section covers the study of effect of droplet shape on its evaporation rate. As discussed previously, the droplet shape is governed by Weber number. In order to analyze the interaction of droplet shape with its evaporation rate, an isolated droplet with various Weber numbers under a convective flow is studied. 

\textcolor{black}{
The droplet is nearly stationary and stays at its predetermined location, for a fixed $U_{rel}$, $Re$ based on initial diameter $d_0$ is constant and $We$ is modified by changing the surface tension. It is also important to recall that these non-dimensional numbers are constant and do not vary with time.}
$Re=25$ and $Re=120$ are selected for this study to cover a range of boundary layer behaviors. At $Re=25$, the flow over a solid sphere is fully attached and at $Re=120$, the flow behind a solid sphere is clearly separated with the steady ring vortex formation \cite{taneda1956}. \textcolor{black}{The range of Weber numbers selected for this study is $We=1-12$ \cite{loth2006} which leads to the droplet shapes from spherical to highly deformed.}

\textcolor{black}{It is worth discussing the fact that as the droplets deform, they move from a spherical shape to a disk-like shape. More importantly, the effective diameter of the disk is larger than that of the original sphere, and this will affect the flow dynamics. In this work, this effect is captured implicitly through the measurement and discussion of the Weber number. However, no explicit measurements of the droplet deformation are made to quantify its effect on the flow. Such an approach is useful in practical model development for droplets since parameters such as $Re$ and $We$ are easy to measure. And this approach was chosen here since such model development is the long-term goal of this work.}

The results for these cases are compared in terms of \textcolor{black}{the normalized total evaporation rate (${\dot{m}}_{ND}$) and its contributors: normalized local evaporation rate per unit area (normalized local evaporation flux) (${\dot{m''}}_{ND} $ ), normalized surface area ($A/A_0$).}
\textcolor{black}{These parameters are normalized using the total evaporation rate of droplet under quiescent flow based on $d^2$ law \cite{turns2000}, defined as ${\dot{m}}_{d^2} = 4 \pi r_0 \rho D ln(1+ B)$.}

\subsubsection{Normalized Total Evaporation Rate}

\begin{figure}[H]
   \centering
   \includegraphics[width=0.7\linewidth]{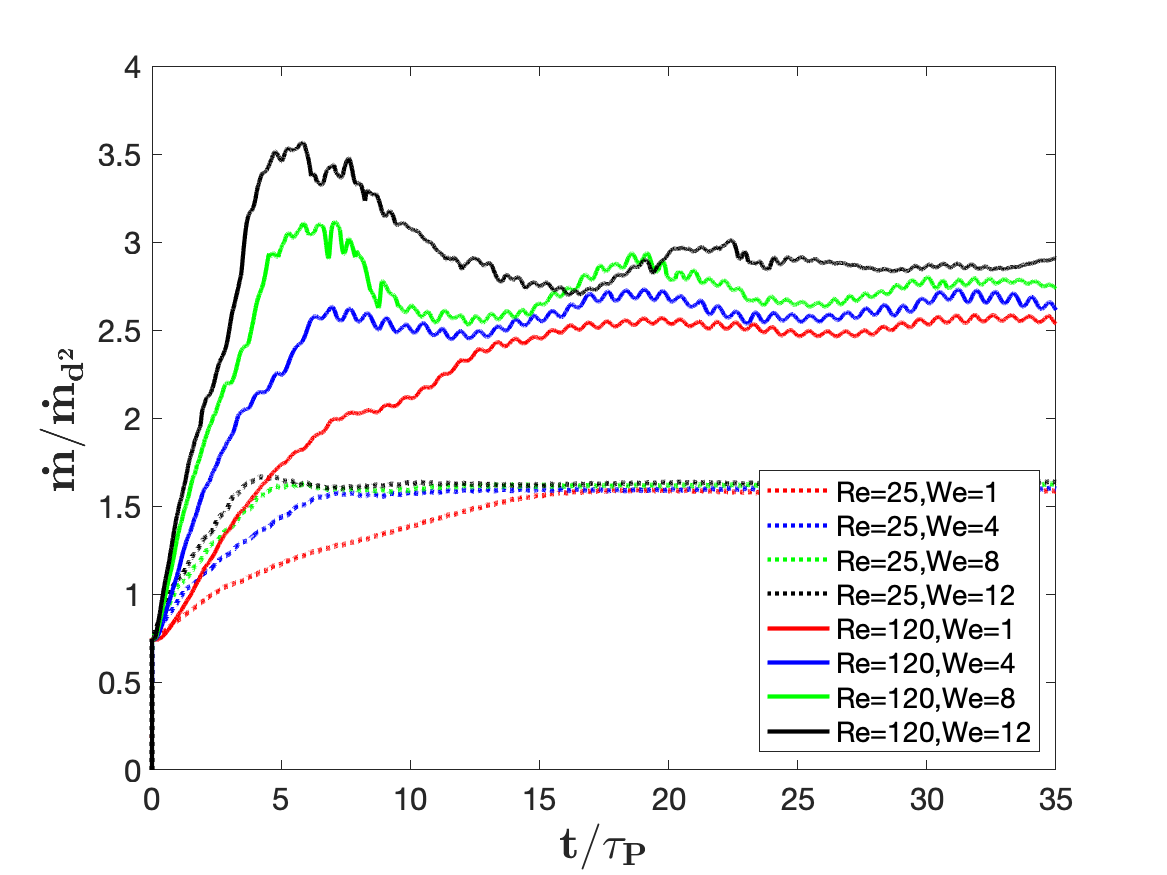} 
   \caption{Normalized total evaporation rate (${\dot{m}}_{ND}$ ) vs normalized time ($t/\tau_P$) at $Re=25$ and $Re=120$}
   \label{fig:Ch2_total_mdot}
\end{figure}

\begin{figure}[h!]
   \centering
   \includegraphics[width= 0.7\linewidth]{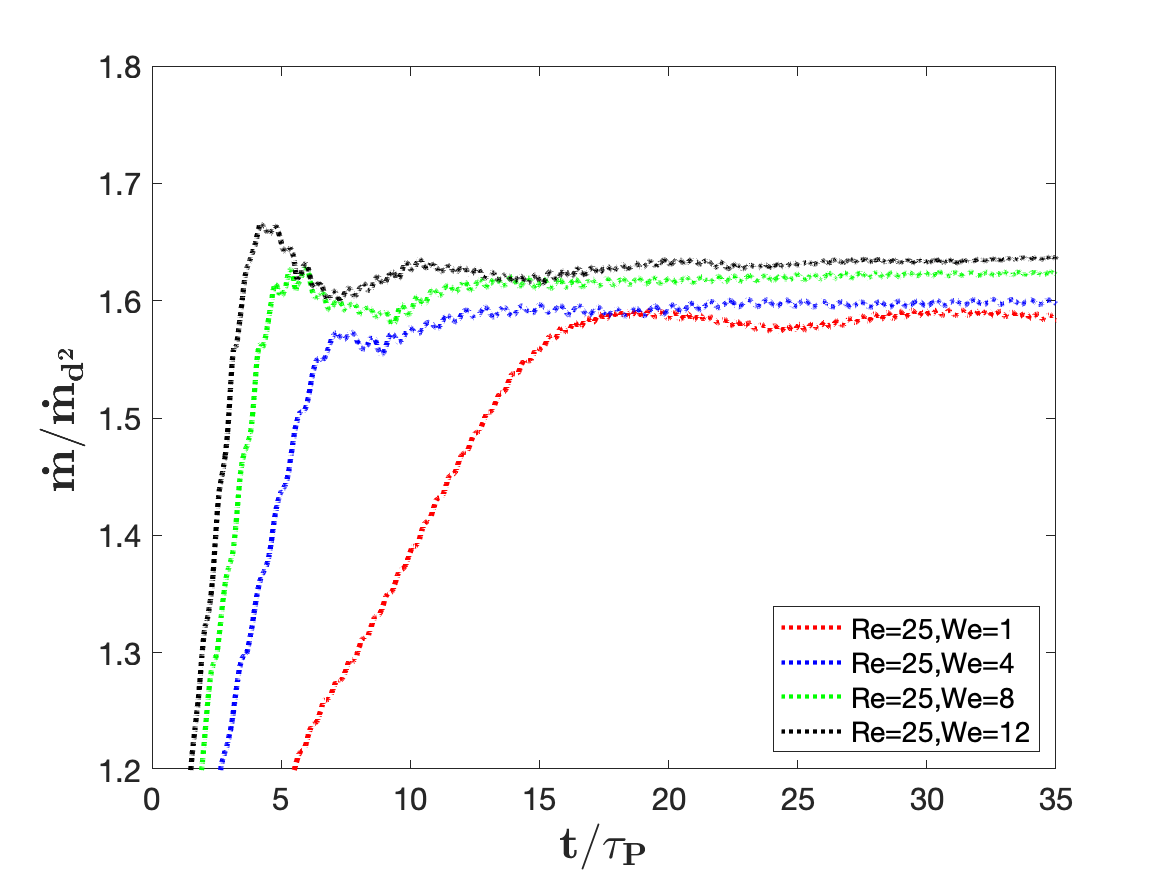} 
   \caption{Zoomed-in plot for ${\dot{m}}_{ND}$ vs normalized time at $Re=25$}
   \label{fig:Ch2_total_mdot_zoomedin}
\end{figure}

\Cref{fig:Ch2_total_mdot} shows the time history of normalized total evaporation rate (${\dot{m}}_{ND}$). In all cases, the instantaneous evaporation rate starts at zero and rapidly increases. Since the droplet is preheated to the saturation temperature, this transient period is due to the adjustment of the gaseous flow field around the droplet. After some period of time, a plateau-like behavior is observed. At low Reynolds number $Re=25$, the difference in ${\dot{m}}_{ND}$ is seen in the beginning for $We=1-12$ (marked in dotted lines). However, at later time ($t/\tau_P >15$), ${\dot{m}}_{ND}$ reaches a steady value for all Weber numbers. 
A more closer (zoomed-in) look at the results at $Re=25$ displayed in  \cref{fig:Ch2_total_mdot_zoomedin} shows an increasing trend in ${\dot{m}}_{ND}$  with increase in $We$. Based on the spatially and time-averaged solution (after $t/\tau_P >15$), the $\%$ increase in ${\dot{m}}_{ND}$  at $We=12$ with respect to $We=1$ is $~3\%$.
At $Re=120$, a significant increase in total evaporation rate is observed with increase in Weber number as shown in the  \cref{fig:Ch2_total_mdot}. This suggests that the Weber number has a clear effect on ${\dot{m}}_{ND}$  at higher $Re$. The $\%$ increase in ${\dot{m}}_{ND}$  at $We=12$ with respect to $We=1$ is $~20 \%$. \textcolor{black}{Moreover, an oscillatory trend is observed for all the Weber numbers higher than 1 at both $Re=25,\ 120$ especially during the transient period (till $t/\tau_p <7$). This is more prominent at $Re=120$. This appears to be due to  oscillations in droplet surface area in \cref{fig:Ch2_area}.}

\subsubsection{Normalized Local Evaporation Flux and Normalized Total Surface Area}

\begin{figure}[h!]
   \centering
   \includegraphics[width=0.7\linewidth]{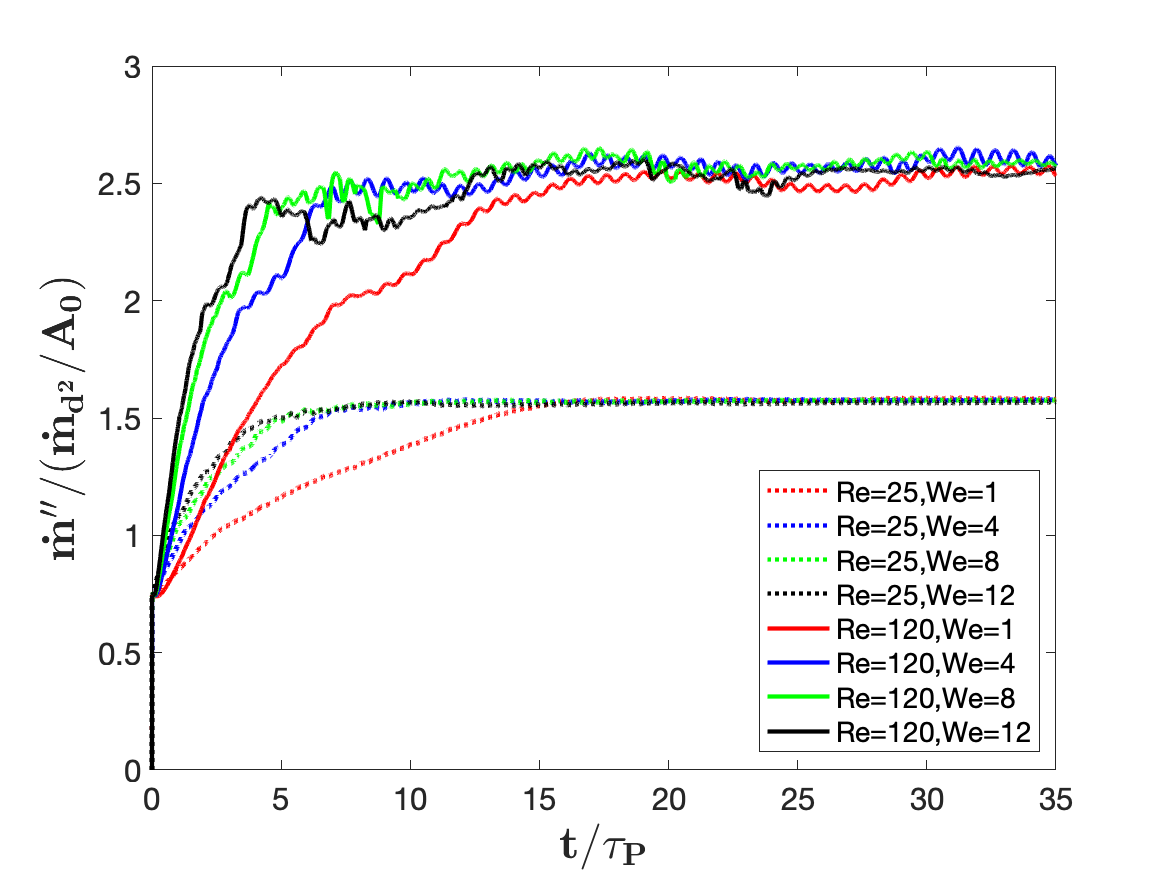} 
   \caption{Normalized local evaporation flux (${\dot{m''}}_{ND}$) vs non-dimensional time ($t/\tau_P$) at $Re=25$ and $Re=120$}
   \label{fig:Ch2_local_mdot}
\end{figure}

\begin{figure}[h!]
   \centering
   \includegraphics[width=0.7\linewidth]{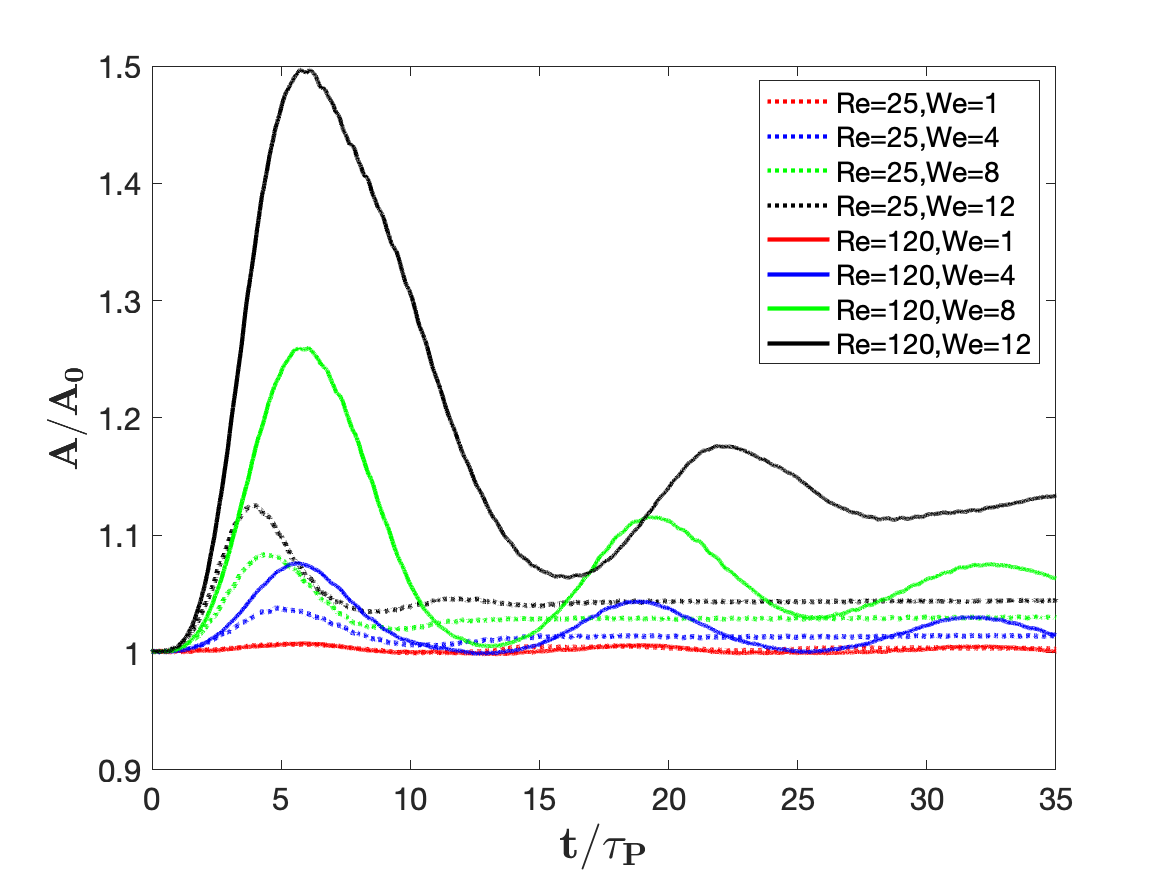} 
   \caption{Non-dimensional Surface Area ($A/A_0$) vs non-dimensional time ($t/\tau_P$) at $Re=25$ and $Re=120$}
   \label{fig:Ch2_area}
\end{figure}

To find out the contribution of local evaporation flux ($\dot{m''}$) and surface area at $Re=25$ and $Re=120$, \textcolor{black}{the results for normalized surface-averaged ${\dot{m''}}_{ND}$} and normalized surface area ${A/A_0}$ are plotted in  \cref{fig:Ch2_local_mdot} and \cref{fig:Ch2_area}, respectively. Results in  \cref{fig:Ch2_local_mdot} show that at $Re=25$, ${\dot{m''}}_{ND}$ is marginally higher ($\sim 1 \%$) for $We=1$ in comparison to $We=12$. 
At $Re=120$, the effect of Weber number is clearly seen on the local evaporation flux at early time. Initially, as Weber number increases, ${\dot{m''}}_{ND}$ increases. However at later time, a clear trend can not be seen.

This trend in ${\dot{m''}}_{ND}$ seems to contradict its relation with curvature as per theoretical analysis by Tonini and Cossali \cite{tonini2013, TONINI2016} and Palmore \cite{palmore_JHT_2022}. These previous studies were performed at $Re \approx 0$.
However, in our studies, due to presence of convective flow over the droplet, the flow is more complex, for example, the flow separation is observed for both $Re=25$ and $Re=120$. In \cref{sec:flow}, we will discuss how the theories by Tonini and Cossalli and Palmore can be used to make limited predictions about the local evaporation fluxes, even for high Reynolds number studies.

\begin{figure}[h!]
   \centering
   \includegraphics[width=4.5cm]{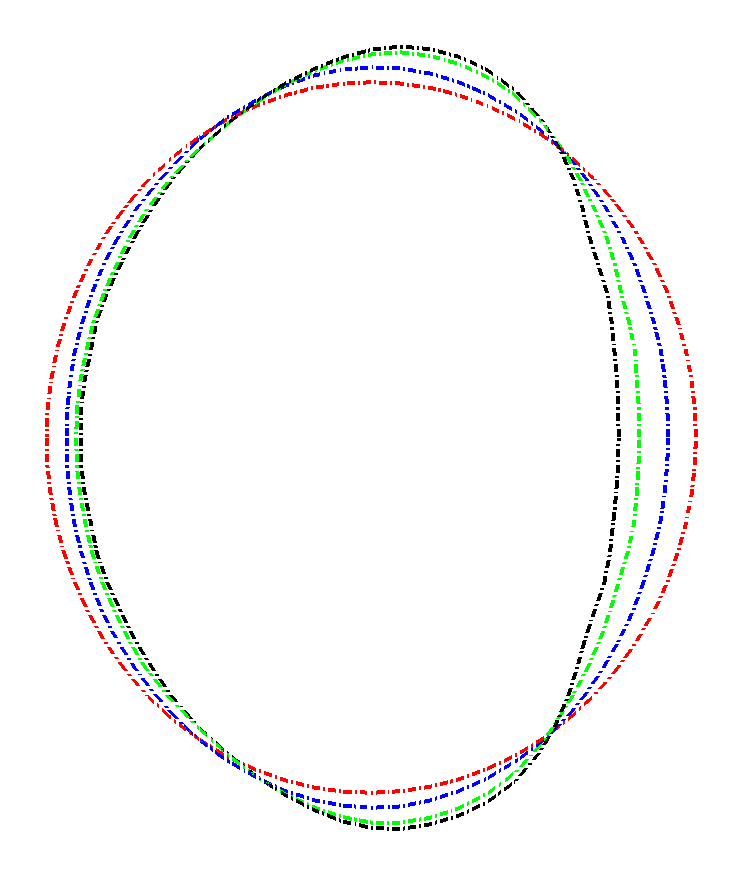} 
   \caption{$Re=25$: Cross sectional view of droplet shape at $t/\tau_P=20$. Red : $We=1$, Blue : $We=4$, Green: $We=8$, Black : $We=12$}
   \label{fig:area_re_25_t_20tcap}
\end{figure}

\Cref{fig:Ch2_area} shows the normalized total surface area ($A/A_0$) evolution with time. Here, $A_0$ is the initial droplet surface area. The results suggest that at any particular Reynolds number, the droplet higher than $We=1$ deforms significantly and reaches a peak of total surface area. \textcolor{black}{Oscillations in droplet shape are reflected in the normalized area plot in \cref{fig:Ch2_area} where two peaks are observed for every Weber number case}. Moreover, this observation about the peaks in surface area ratio plot correlate well with the results shown in DNS study by Schlottke and Weigland \cite{SCHLOTTKE20085215}. 
At $Re=25$, the steady value of maximum normalized area is after $t/\tau_P>20$ is 1.04 and it occurs for $We=12$. 
The cross section area in z-plane at time instance $t/\tau_P =20$ is compared at $Re=25$ in \cref{fig:area_re_25_t_20tcap}. The line style and color in this figure is kept consistent with plots for ${\dot{m}}_{ND}$, ${\dot{m''}}_{ND}$ and $A/A_0$ in  \cref{fig:Ch2_total_mdot} -  \cref{fig:Ch2_area}.

\begin{figure}[h!]
   \centering
   \includegraphics[width=4.5cm]{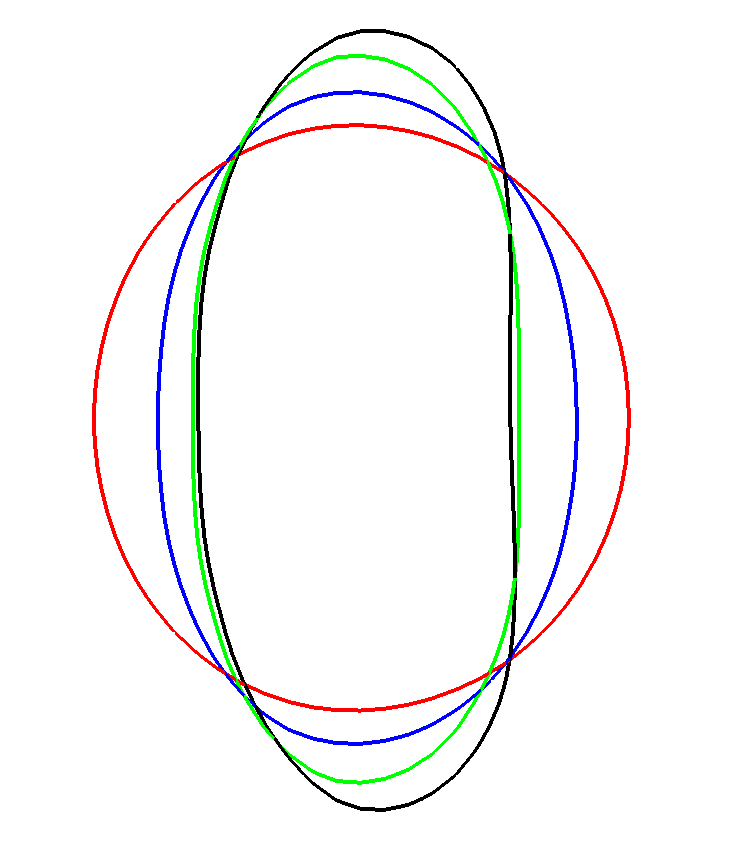} 
   \caption{$Re=120$ : Cross sectional view of droplet shape at $t/\tau_P=20$. Red : $We=1$, Blue : $We=4$, Green: $We=8$, Black : $We=12$}
   \label{fig:area_re_120_t_20tcap}
\end{figure}

At $Re=120$, this increase in area is nearly $~20 \%$ at $t/\tau_P=20$ with respect to a spherical droplet and is highest for $We=12$.
% due to higher deformation.
A sectional view of droplet shape at this time instance is shown in  \cref{fig:area_re_120_t_20tcap}. From this figure,  $~17 \%$ elongation in length perpendicular to flow with respect to initial diameter of spherical droplet is observed for $We=12$.
Hence, the net effect of both ${\dot{m''}_{ND}}$ and total surface area leads to increase in the total evaporation rate in case of $We=12$ as shown  \cref{fig:Ch2_total_mdot}.
% ------------post-processing----------------------------------------------------
\subsubsection{Analysis of local evaporation flux and gas flow} \label{sec:flow}
\begin{figure}[h!]
    \centering
    \begin{subfigure}{0.49\linewidth}
    \centering
    \includegraphics[width=\linewidth]{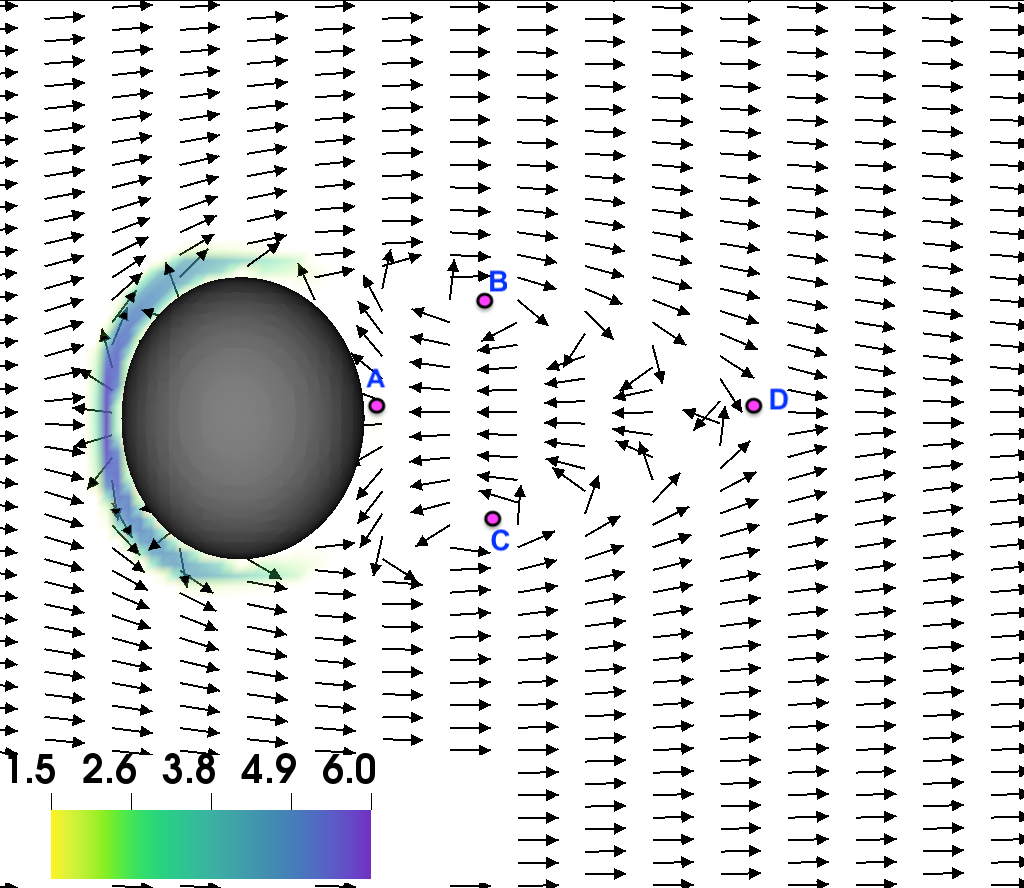}
    \caption{$We=1$}
    \end{subfigure}
    \begin{subfigure}{0.49\linewidth}
    \centering
    \includegraphics[width=\linewidth]{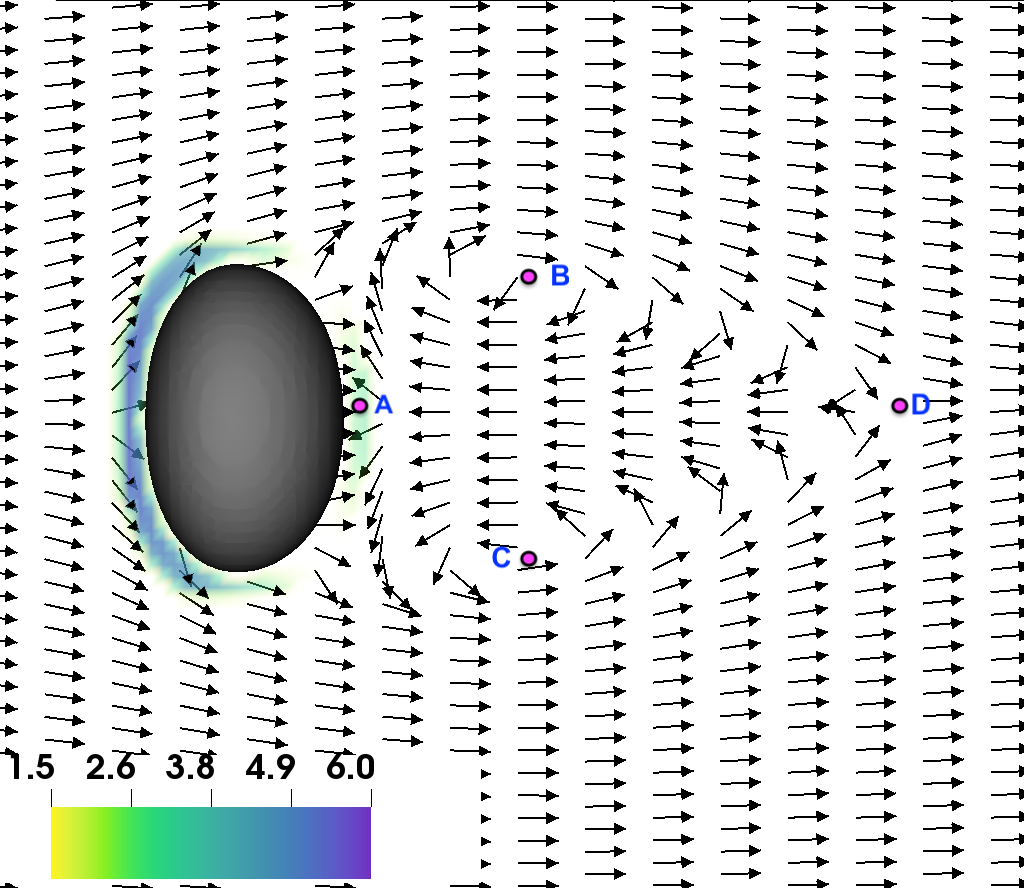}
    \caption{$We=4$}
    \end{subfigure}
    
    \begin{subfigure}{0.49\linewidth}
    \centering
    \includegraphics[width=\linewidth]{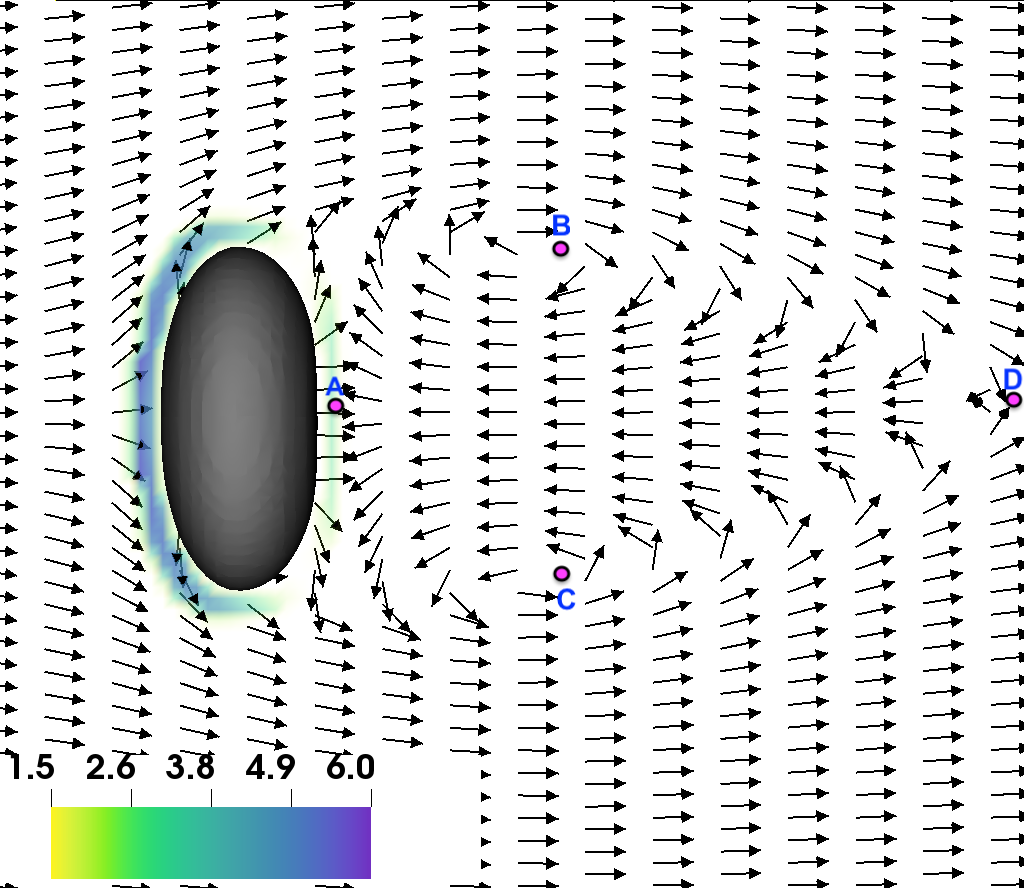}
    \caption{$We=8$}
    \end{subfigure}
    \begin{subfigure}{0.49\linewidth}
    \centering
    \includegraphics[width=\linewidth]{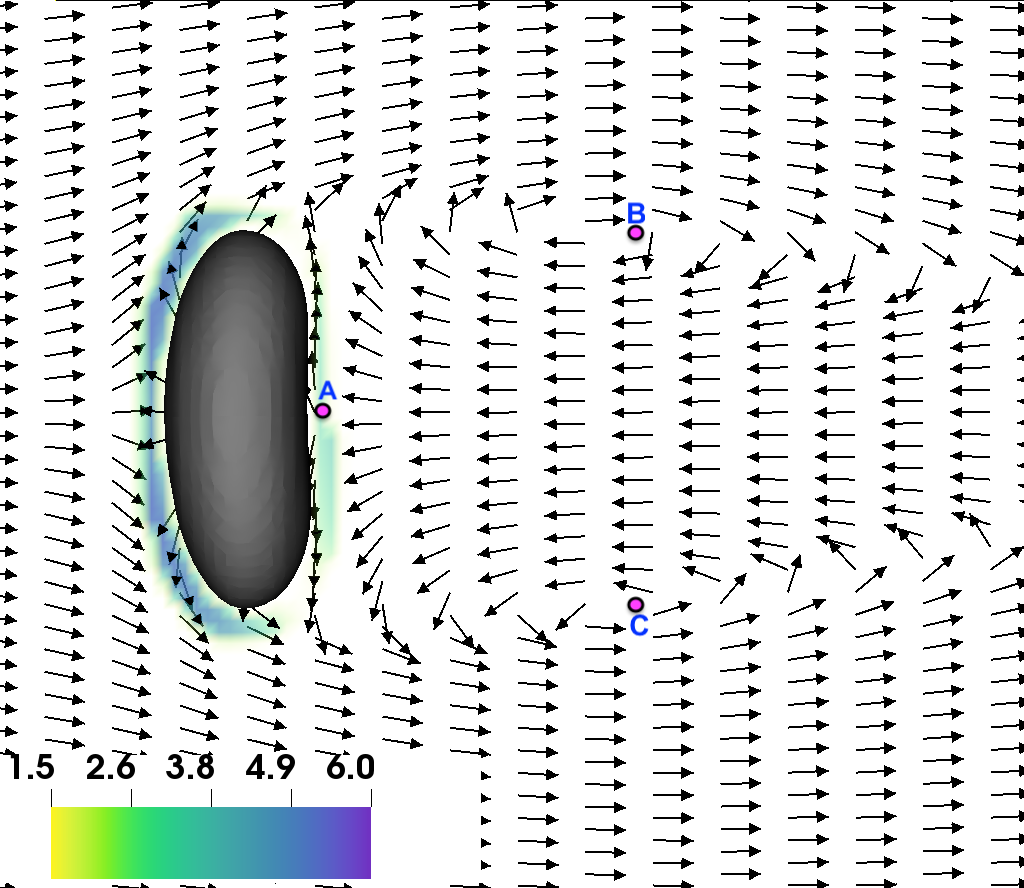}
    \caption{$We=12$}
    \end{subfigure}
\caption{$Re=120 $ : ${\dot{m''}}_{ND}$  for different Weber numbers at $t/\tau_{p} =20$}
\label{fig:flow_Re_120}
\end{figure}

The qualitative analysis of the flow field at time instant $t =20 \tau_P$ is shown in \cref{fig:flow_Re_120}. The psuedocolors of local evaporation flux along with the velocity vectors at $Re=120$ for various Weber numbers are shown. In addition to this, four probes and a line segment are located inside the wake region to record the flow related parameters. Probe A is located near the rear stagnation point, probe B and C are located at the center of vortex and probe D is located at the end of wake.

These results show increasing deformation with increase in Weber number. This deformation correlates with a larger wake, both in terms of wake height $H$(measured between points B and C) and wake length $L$ (measured between points A and D). \textcolor{black}{Considering the flow axisymmetry, the approximate non-dimensional volume of wake scales as $ V^* \sim (H/(2d))^2 \cdot (L/d)$}. This data is tabulated in \cref{table:2}. 
\begin{table} [h!]
\begin{center}
\begin{tabular}{ |p{1cm}|p{3cm}|p{2cm}|p{2cm}|p{2cm}|} 
\hline
\bm{$We$} & \bm{${\dot{m}^{''}}_{ND}$} at Pt. A  &  \bm{$L/d$}& \bm{$H/d$} & \bm{$V^*$} \\ 
\hline
\hline
1 & 2.37  & 1.89 &  0.87&  0.36 \\
\hline
4 & 3.06  & 1.99 &  1.13& 0.64\\
\hline
8 & 3.34 & 2.45 &  1.19 & 0.88\\ 
\hline
12 & 3.62 & 3.42 &  1.45 & 1.86 \\ 
\hline
\end{tabular}
\end{center}
\caption{$Re=120$: Measurements at different points inside wake region}
\label{table:2}
\end{table}

\begin{figure}[h!]
    \centering
    \begin{subfigure}{0.49\linewidth}
    \centering
    \includegraphics[width=\linewidth]{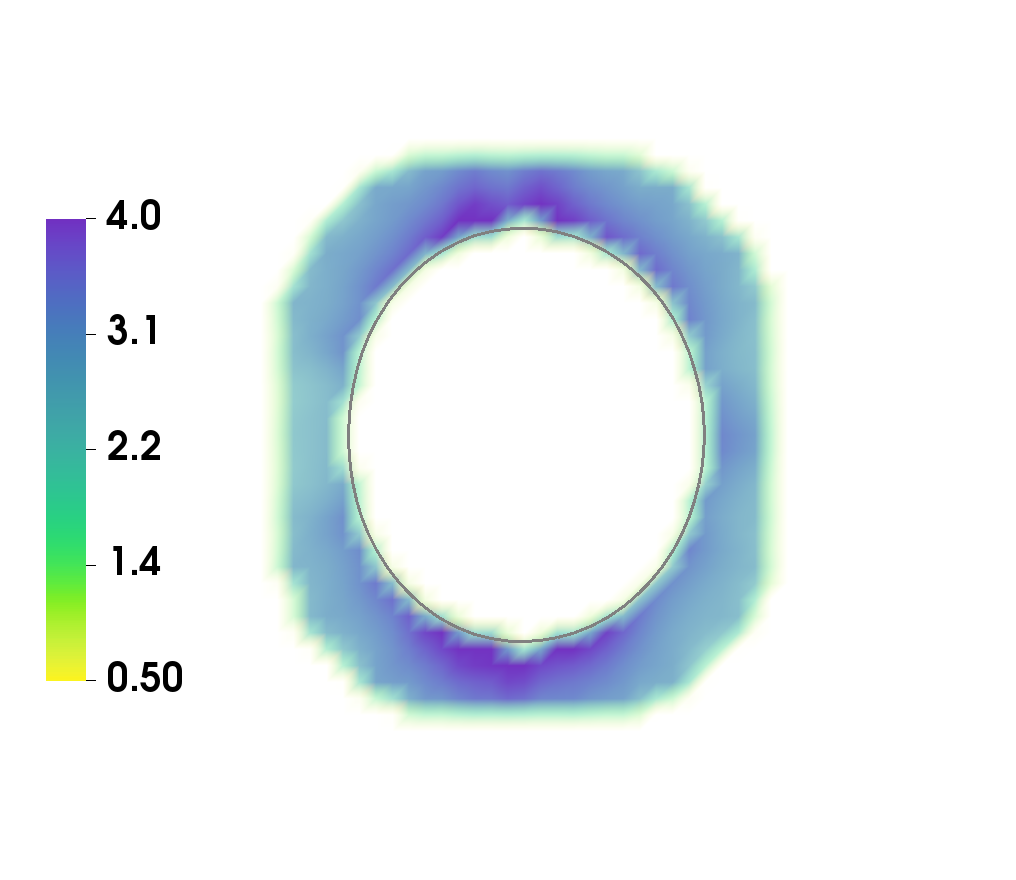}
    \caption{$We=1$: Normalized curvature}
    \label{fig:curv_we_1}
    \end{subfigure}
    \begin{subfigure}{0.49\linewidth}
    \centering
    \includegraphics[width=\linewidth]{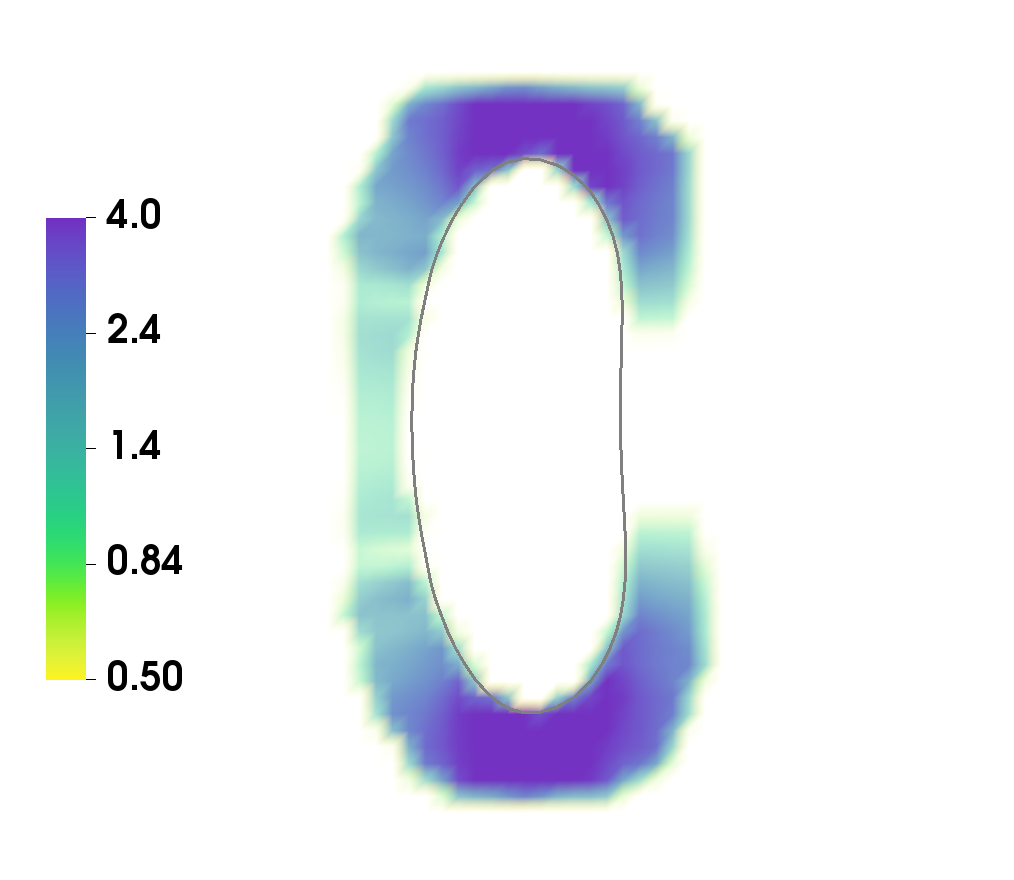}
    \caption{$We=12$ : Normalized curvature}
    \label{fig:curv_we_12}
    \end{subfigure}
    
    \begin{subfigure}{0.49\linewidth}
    \centering
    \includegraphics[width=\linewidth]{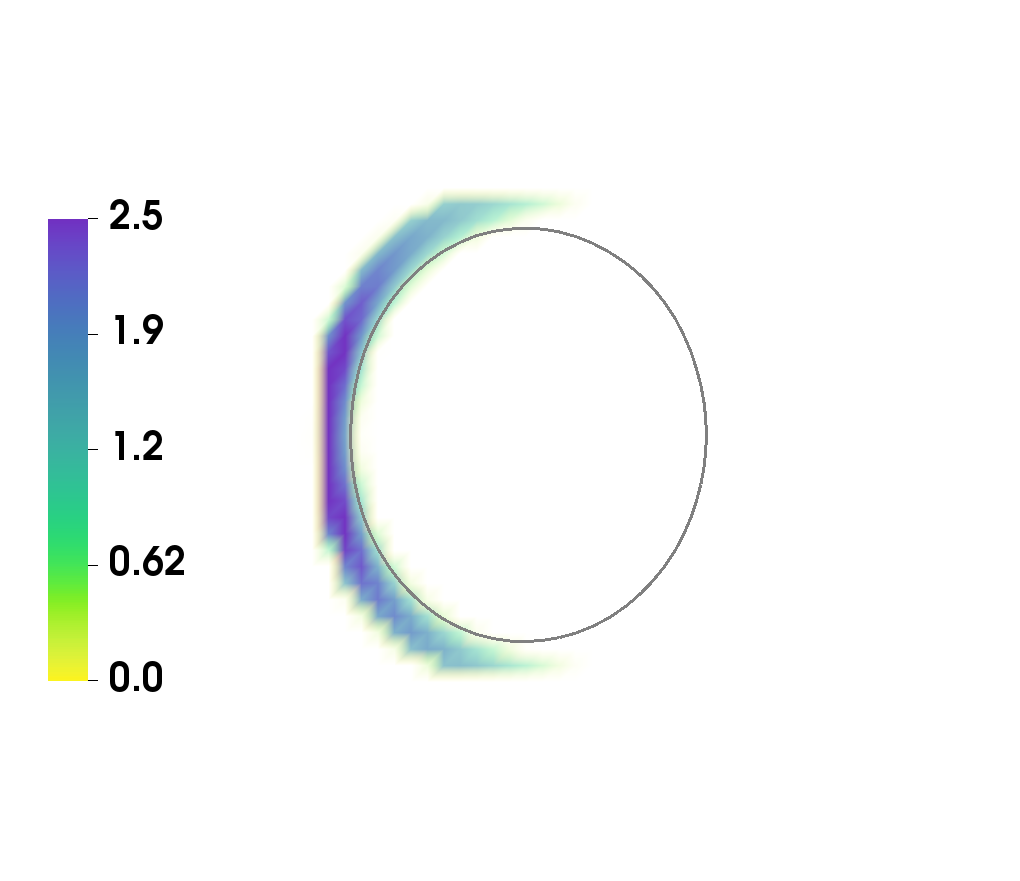}
    \caption{$We=1$: $\Delta{\dot{m''}_{ND}}$}
    \label{fig:vof_mdot_we_1}
    \end{subfigure}
    \begin{subfigure}{0.49\linewidth}
    \centering
    \includegraphics[width=\linewidth]{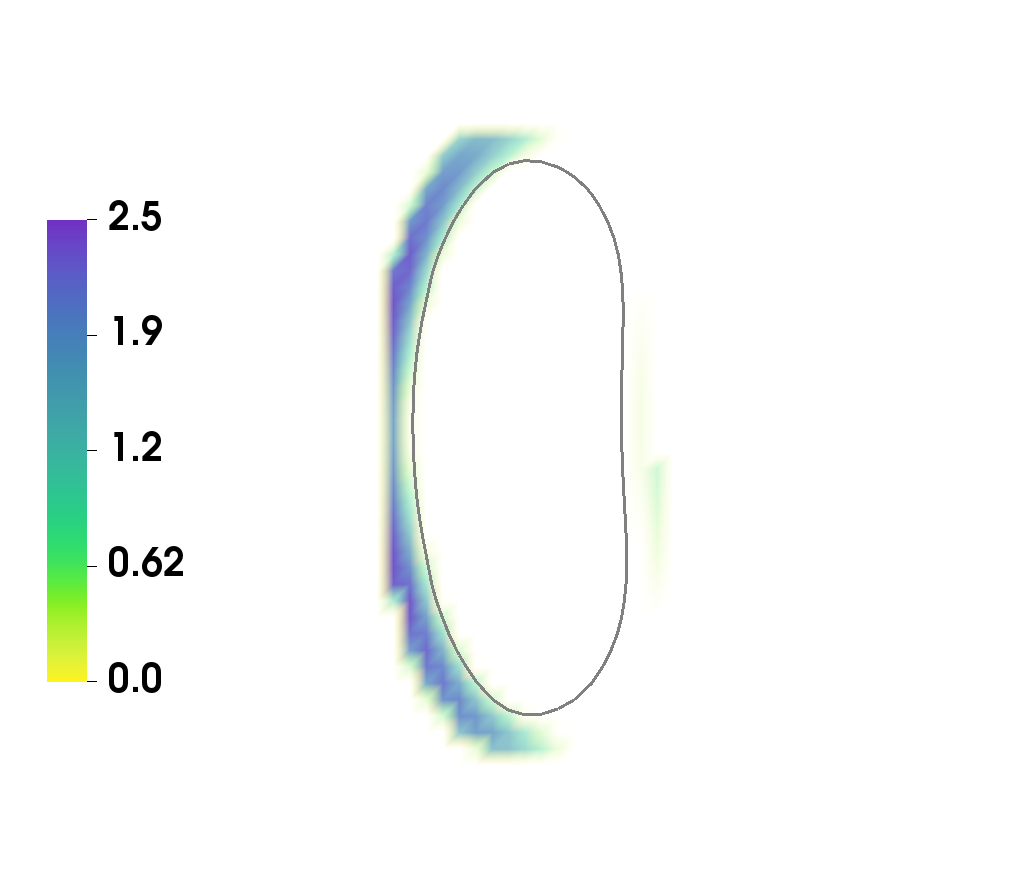}
    \caption{$We=12$: $\Delta{\dot{m''}_{ND}}$}
    \label{fig:vof_mdot_we_12}
    \end{subfigure}
\caption{\textcolor{black}{$Re=120 $ : Distribution of normalized curvature and normalized $\Delta \dot{m}''$ at $t/\tau_{P} =20$}}
\label{fig:curv_re_120_we}
\end{figure}

In terms of local evaporation flux, the majority of the evaporation occurs at the front of the droplet (i.e. the windward side). Moreover, the evaporation flux in this region seems dependent on the local curvature of droplet as shown in \cref{fig:curv_re_120_we}. \textcolor{black}{To elaborate on this, the psuedocolors of normalized curvature and $\Delta {\dot{m''}}_{ND}$ are plotted in \cref{fig:curv_re_120_we}. The curvature is normalized using curvature of a sphere of diameter $d_0$} and $\Delta {\dot{m''}}_{ND}$ is defined as $(\dot{m''} - \dot{m''}_{avg})/(m_0/A_0 \tau_D)$, where $\dot{m''}_{avg}$ averaged over time. 
Comparing $We=1$ with $We=12$, the curvature is higher in the front region which corresponds to the area of higher local $\dot{m''}_{ND}$.

Given this insight, the results of this study can now be compared with the low $Re$ theories for evaporation by Tonini and Cossali \cite{tonini2013, TONINI2016}, and Palmore \cite{palmore_JHT_2022} \textcolor{black}{which suggest the direct dependence of evaporation flux on the surface curvature. At the upstream stagnation point, the flow speed approaches zero, and accordingly the low Re theory is appropriate. Hence, the evaporation flux is found to be proportional to the curvature (shown in \cref{fig:curv_we_1} and \cref{fig:vof_mdot_we_1}) as predicted by theory given by \cite{tonini2013,TONINI2016,palmore_JHT_2022}}. However, as the flow develops around the droplet, this low $Re$ approximation becomes inappropriate, and the curvature can no longer accurately predict evaporation flux. Furthermore, the low $Re$ theories do not make any predictions for boundary layer separation, hence the previous theories have no validity for flow dynamics beyond the separation point. 

It is also interesting to note that, for the non-spherical droplet such as \cref{fig:curv_we_12}, the region of highest curvature occurs at the top and bottom of the droplet near the separation point, whereas the front stagnation region has a region of lower curvature than the corresponding sphere. Hence, the curvature-induced vaporization enhancement predicted by low Re theories is not experienced in practice for these highly deformed droplets. The net effect of these two phenomena imply that the evaporation flux for the deformed droplet on the front of the droplet must be lower than that of the spherical droplet. In addition to this, it will be demonstrated later that the evaporation flux on the rear is higher \textcolor{black}{for deformed droplets}, resulting in approximately equal overall local evaporation flux. This effect is also visible in the rear in \cref{fig:vof_mdot_we_12}.

\begin{figure}[h!]
    \centering
    \begin{subfigure}{0.7\linewidth}
    \centering
    \includegraphics[width=\linewidth]{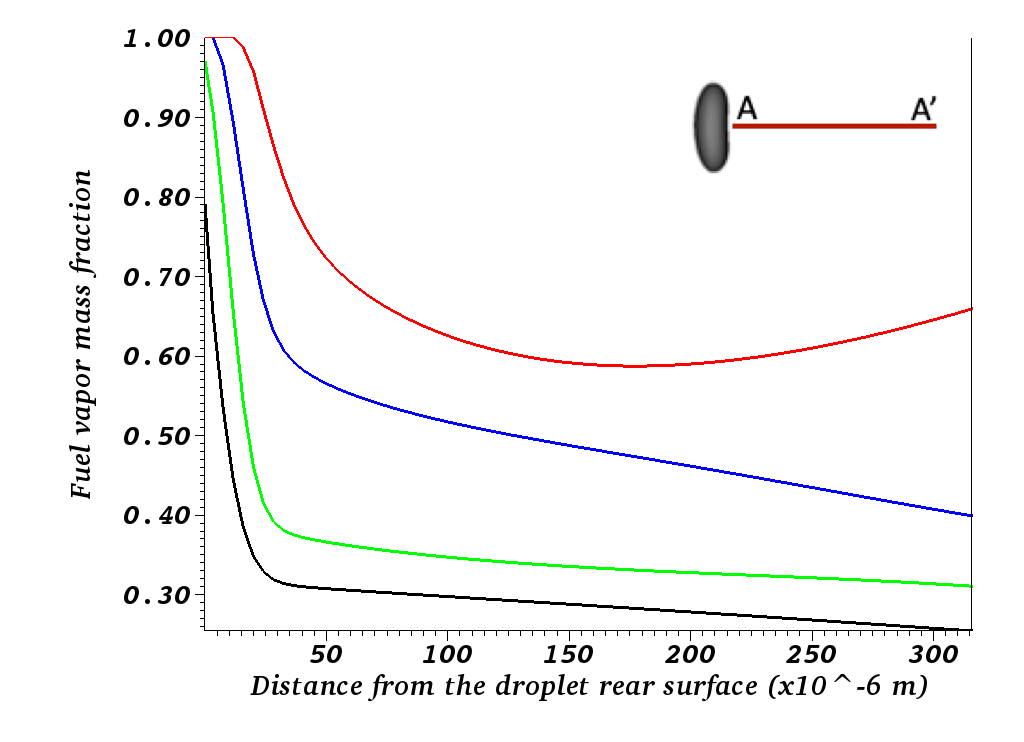}
    \caption{Vapor Mass Fraction}
    \label{fig:lineout_gy}
    \end{subfigure}
    \begin{subfigure}{0.7\linewidth}
    \centering
    \includegraphics[width=\linewidth]{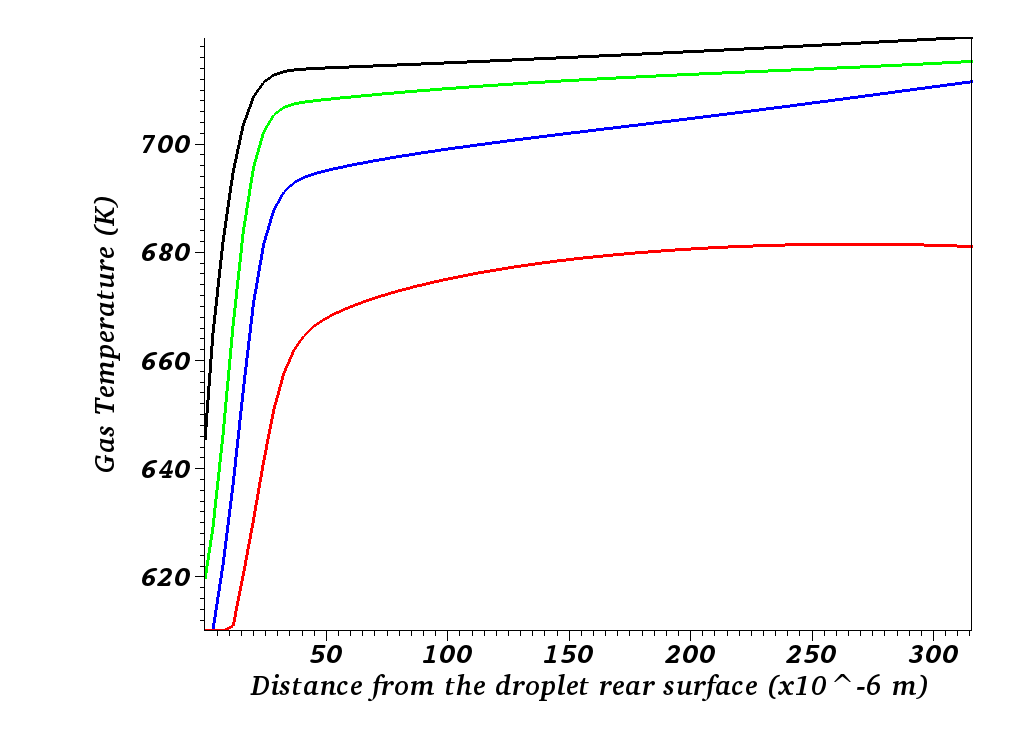}
    \caption{Gas Temperature}
    \label{fig:lineout_gt}
    \end{subfigure}
\caption{$Re=120 $: Vapor Mass Fraction and Gas Temperature along the line segment $AA'$.
        Red : $We=1$, Blue : $We=4$, Green: $We=8$, Black : $We=12$}
\label{fig:lineout}
\end{figure}

This enhancement in the downstream region can be attributed to the wake interaction with the fuel vapor. The velocity vectors in  \cref{fig:flow_Re_120} show the presence of a vortex ring formation for all cases. The length of wake region is observed to be increasing with increase in Weber number. These details are quantified and tabulated in \cref{table:2}. It is clear that scaled volume of wake region is nearly $5$ times larger for $We=12$ in comparison to $We=1$. As vapor mixing occurs only in wake, the larger wake is leading to the lower average vapor concentration. Since evaporation is driven by the gradient of vapor concentration between the liquid surface and its surrounding, the evaporation flux in this region is enhanced. An analogous claim can be made for the temperature, however in this case the temperature is increased in the wake due to mixing. This viewpoint is supported by~\cref{fig:lineout}, which delineates the vapor mass fraction and temperature profiles behind the droplet. The data are extracted on a line segment that starts at ($-0.7d_0,0,0$) (near the rear of the droplet) and ends at ($2.5d_0,0,0$) to cover the maximum wake length, as demonstrated in~\cref{fig:lineout_gy}. The vapor mass fraction is decreasing monotonically with $We$, whereas the temperature has the opposite trend. In both cases, the trend indicates mixing of values between the interface conditions and the farfield conditions. \textcolor{black}{Accordingly, this leads to higher gradients and higher evaporation rates in the wake region as shown in \cref{fig:flow_Re_120} with increase in $We$ number.} 

%-------------------------------------------------------------
\section{Conclusions}\label{sec:conclustion}

With the motivation to understand the complex relationship between the droplet shape and its evaporation rate, this work covers the droplet evaporation under convective flow using interface capturing Direct Numerical Simulation for multiphase flows. The liquid-gas interface can deform freely in these studies. A grid independence study is performed and the solver accuracy is validated against empirical correlations for droplet evaporation by Abramzon and Sirignano \cite{abramzon1989}. 

% \textcolor{black}{In case of pure evaporation of droplet, due to continuous mass loss from the droplet surface, the liquid volume as well as the surface area reduce in time. Therefore, the cumulative effect of evaporation flux and surface area on total evaporation rate can be ambiguous. To separate these two effects, a well-known ``Quasi-steady evaporation'' approach is implemented where the liquid volume is artificially kept constant in simulations while still solving the remaining equations for the fluid dynamical system.} 
% \textcolor{black}{For a constant relative inflow velocity, the droplet shape is varied by modifying Weber number from $We=1-12$}. The convective flow velocity is modified by changing initial Reynolds number $Re=25,120$.

\textcolor{black}{
% In case of pure evaporation of droplet, due to continuous mass loss from the droplet surface, the liquid volume as well as the surface area reduce in time. Therefore, the cumulative effect of evaporation flux and surface area on total evaporation rate can be ambiguous.
To analyze the effect of droplet shape on its evaporation rate, quasi-steady approach is implemented, where the liquid volume is artificially kept constant in simulations while still solving the remaining equations for the fluid dynamical system.} 
\textcolor{black}{For a constant relative inflow velocity, the droplet shape is varied by modifying Weber number from $We=1-12$}. The convective flow velocity is modified by changing initial Reynolds number $Re=25,120$. 

At $Re=25$, the increase in total evaporation rate at $We=12$ with respect to $We=1$ is $~3\%$ which is marginal. However, at $Re=120$ and $We=12$,  $20 \%$ enhancement in total evaporation rate is observed when compared to $We=1$. To understand the source of this enhancement, the total evaporation rate was decomposed into evaporation per unit surface area ($\dot{m''}$) and surface area.
This enhancement is mostly coming from increase in surface area due to deformation and the averaged value of evaporation flux seems to be nearly same for all Weber number.

At first glance, this appears to contradict the established theories for droplet evaporation Tonini and Cossali \cite{tonini2013, TONINI2016}, Palmore \cite{palmore_JHT_2022}. However, this effect is due entirely due to the high Reynolds number phenomena including boundary layer separation. Interestingly, for the front (i.e. windward) stagnation region that forms over the droplet, the evaporation flux is proportional to the interface curvature, as predicted by the low Re theory. However, as a consequence of the boundary layer behavior, the evaporation flux for deformed droplets is lower on their fronts when compared to nearly spherical droplets.

After the flow separation, the formation of wake and its interaction with fuel vapor is observed at the downstream. Due to this interaction, the evaporation flux on the rear (i.e. leeward) side of the droplet is enhanced where the flow is separated. This effect largely cancels out the diminished evaporation flux on the front of the droplets, and results in similar values for surface averaged evaporation flux. This emphasizes the importance of boundary layer development in evaporation for higher Reynolds number flow and suggests a more complex relationship between Weber Number, Reynolds Number, and evaporation phenomena than had been previously predicted.

%-------------------------------------------------------------
\section{Acknowledgement}
The authors acknowledge Advanced Research Computing at Virginia Tech for providing computational resources and technical support that have contributed to the results reported in this paper.



 \bibliographystyle{elsarticle-num} 
 \bibliography{mybibfile.bib}





% \end{thebibliography}
\end{document}